\documentclass[submission,copyright,creativecommons,sharealike,noncommercial]{eptcs}
\providecommand{\event}{QPL 2016}
\pdfoutput=1

\usepackage{mathtools} % Loads and extends amsmath
\usepackage{amssymb} % Extram mathematical symbols (loads amsfonts)
\usepackage{bbold} % Sans-serif blackboard bold font for numbers
\usepackage{amsthm} % Theorem environments
\usepackage{relsize} % Additional relative sizes for fonts
\usepackage{microtype} % Improves appearance of writing
\usepackage{hyperref} % Hyperlink citations. Comment out for arxiv submission
\usepackage[nocompress]{cite} % Comment out if using natbib 
\usepackage{graphicx} % To import graphics
\usepackage[usenames,dvipsnames]{xcolor} % Introduces colour names
\usepackage{tikz} % TikZ
\usepackage{tikzfig} % Additional TikZ macros for categorical quantum mechanics
\usepackage{circuitikz} % TikZ circuit diagrams 
\usetikzlibrary{
	arrows,
	shapes,
	decorations,
	intersections,
	backgrounds,
	positioning,
	circuits.ee.IEC
	}
		\newcounter{theorem_c} % Unified coutner for all theorem environments...
		\numberwithin{theorem_c}{section} % ... numbered within sections
		\numberwithin{equation}{section} % Equations are also numbered within sections (but have a separate counter)
		\theoremstyle{plain} 
		\newtheorem{theorem}[theorem_c]{Theorem}
		
		\newtheorem{lemma}[theorem_c]{Lemma}

		\theoremstyle{definition}

	\newcommand{\emptyArg}{\,\underline{\hspace{6px}}\,} % Underscore for omitted, anonymous arguments
	\newcommand{\inlineQuote}[1]{\textquotedblleft #1\textquotedblright} % Left-right quotes surrounding #1
	\newcommand{\naturals}{\mathbb{N}} % Set of natural numbers
	\newcommand{\integers}{\mathbb{Z}} % Set of interer numbers
	\newcommand{\rationals}{\mathbb{Q}} % Set of rational numbers
	\newcommand{\reals}{\mathbb{R}} % Set of real numbers
	\newcommand{\complexs}{\mathbb{C}} % Set of complex numbers
	\newcommand{\integersMod}[1]{\mathbb{Z}_{#1}} % Set/group/ring of integers mod #1
	\newcommand{\modclass}[2]{#1 \; (\text{mod } #2)} % Equivalence class of integers mod #2 corresponding to representative #1
	\newcommand{\nonstd}[1]{\,^\star #1}
	\newcommand{\starNaturals}{\nonstd{\naturals}} % Set of non-standard naturals
	\newcommand{\starIntegers}{\nonstd{\integers}} % Set of non-standard integers
	\newcommand{\starRationals}{\nonstd{\rationals}} % Set of non-standard rationals
	\newcommand{\starComplexs}{\nonstd{\complexs}} % Set of non-standard complex numbers
	\newcommand{\starReals}{\nonstd{\reals}} % Set of non-standard realsnumbers

	\newcommand{\iffdef}{\stackrel{def}{\iff}} % If and only if (by definition)
	\newcommand{\suchthat}[2]{\left\{#1 \: \middle\vert \: #2\right\}} % Set of elements #1 such that condition #2 holds 
		\newcommand{\ket}[1]{\vert #1 \rangle} % Ket labelled #1
		\newcommand{\bra}[1]{\langle #1 \vert} % Bra labelled #1
		\newcommand{\braket}[2]{\langle #1 \vert #2 \rangle} % Inner product of bra labelled #1 with ket labelled #2
		\newcommand{\innerprod}[2]{\left( #1 , #2 \right)}
		\newcommand{\Trace}[1]{\operatorname{Tr}[#1]} % Trace of an operator
		\newcommand{\pTrace}[4]{\operatorname{Tr}_{#2,#3}^{#4}[#1]} % Trace of an operator
		\newcommand{\LtwoSym}{\operatorname{L}^2} % Symbol for L2 spaces
		\newcommand{\Ltwo}[1]{\LtwoSym[#1]} % L2 space over space #1
		\newcommand{\ltwoSym}{\ell^2} % Symbol for l2 spaces
		\newcommand{\ltwo}[1]{\ltwoSym[#1]} % l2 space over space #1

		\newcommand{\SpaceH}{\mathcal{H}} 
		\newcommand{\SpaceG}{\mathcal{G}}
		\newcommand{\SpaceK}{\mathcal{K}}

		\newcommand{\isom}{\cong} % Isomorphism
		\newcommand{\id}[1]{id_{#1}} % Identity morphism of object #1
		\newcommand{\Hom}[3]{\operatorname{Hom}_{\,#1}\left[#2,#3\right]} % Set of morphisms in category #1 from object #2 to object #3
		\newcommand{\HilbCategory}{\operatorname{Hilb}} % Category of Hilbert spaces
		\newcommand{\sHilbCategory}{\operatorname{sHilb}} % Category of Hilbert spaces
		\newcommand{\fdHilbCategory}{\operatorname{fdHilb}} % Category of finite-dimensional Hilbert spaces
		\newcommand{\starHilbCategory}{^\star\!\HilbCategory} % Category of omega-truncated non-standard hilbert spaces

		\newcommand{\CPMCategory}[1]{\operatorname{CPM}[#1]} % Selinger's CPM category
	\newcommand{\Dcolour}{black}
	\newcommand{\Xbwcolour}{black}
	\newcommand{\Zbwcolour}{white}
	\tikzset{->-/.style={decoration={markings,mark=at position #1 with {\arrow{>}}},postaction={decorate}}}
	\tikzset{-<-/.style={decoration={markings,mark=at position #1 with {\arrow{<}}},postaction={decorate}}}

\tikzstyle{env}=[copoint,regular polygon rotate=0,minimum width=0.2cm, fill=black]

\tikzstyle{probs}=[shape=semicircle,fill=white,draw=black,shape border rotate=180,minimum width=1.2cm]

\tikzstyle{every picture}=[baseline=-0.25em,scale=0.5]
\tikzstyle{dotpic}=[] % for backwards-compatibility
\tikzstyle{diredges}=[every to/.style={diredge}]
\tikzstyle{math matrix}=[matrix of math nodes,left delimiter=(,right delimiter=),inner sep=2pt,column sep=1em,row sep=0.5em,nodes={inner sep=0pt},text height=1.5ex, text depth=0.25ex]

\tikzstyle{inline text}=[text height=1.5ex, text depth=0.25ex,yshift=0.5mm]
\tikzstyle{label}=[font=\footnotesize,text height=1.5ex, text depth=0.25ex,yshift=0.5mm]
\tikzstyle{left label}=[label,anchor=east,xshift=1.5mm]
\tikzstyle{right label}=[label,anchor=west,xshift=-1.5mm]

\tikzstyle{braceedge}=[decorate,decoration={brace,amplitude=2mm,raise=-1mm}]
\tikzstyle{small braceedge}=[decorate,decoration={brace,amplitude=1mm,raise=-1mm}]

\tikzstyle{doubled}=[line width=1.6pt] % set the line width for all doubled (quantum) maps/wires
\tikzstyle{boldedge}=[doubled,shorten <=-0.17mm,shorten >=-0.17mm]
\tikzstyle{boldedgegray}=[doubled,gray,shorten <=-0.17mm,shorten >=-0.17mm]

\tikzstyle{semidoubled}=[line width=1.4pt] % set the line width for all doubled (quantum) maps/wires
\tikzstyle{semiboldedgegray}=[semidoubled,gray,shorten <=-0.17mm,shorten >=-0.17mm]

\tikzstyle{boldedgedashed}=[very thick,dashed,shorten <=-0.17mm,shorten >=-0.17mm]
\tikzstyle{vboldedgedashed}=[doubled,dashed,shorten <=-0.17mm,shorten >=-0.17mm]
\tikzstyle{left hook arrow}=[left hook-latex]
\tikzstyle{right hook arrow}=[right hook-latex]
\tikzstyle{sembracket}=[line width=0.5pt,shorten <=-0.07mm,shorten >=-0.07mm]

\tikzstyle{causal edge}=[->,thick,gray]
\tikzstyle{causal nondir}=[thick,gray]
\tikzstyle{timeline}=[thick,gray, dashed]

\tikzstyle{cedge}=[<->,thick,gray!70!white]

\tikzstyle{empty diagram}=[draw=gray!40!white,dashed,shape=rectangle,minimum width=1cm,minimum height=1cm]
\tikzstyle{empty diagram small}=[draw=gray!50!white,dashed,shape=rectangle,minimum width=0.6cm,minimum height=0.5cm]

\tikzstyle{dot}=[inner sep=0mm,minimum width=3mm,minimum height=3mm,draw,shape=circle,text depth=-0.1mm]
\tikzstyle{ddot}=[inner sep=0mm, doubled, minimum width=3.5mm,minimum height=3.5mm,draw,shape=circle]

\tikzstyle{black dot}=[dot,fill=black]
\tikzstyle{white dot}=[dot,fill=white,,text depth=-0.2mm]
\tikzstyle{green dot}=[white dot] % for backwards-compatibility
\tikzstyle{gray dot}=[dot,fill=gray!40!white,,text depth=-0.2mm]
\tikzstyle{red dot}=[gray dot] % for backwards-compatibility

\tikzstyle{black ddot}=[ddot,fill=black]
\tikzstyle{white ddot}=[ddot,fill=white]
\tikzstyle{gray ddot}=[ddot,fill=gray!40!white]

\tikzstyle{gray edge}=[gray!40!white]

\tikzstyle{small dot}=[inner sep=0.5mm,minimum width=0pt,minimum height=0pt,draw,shape=circle]

\tikzstyle{small black dot}=[small dot,fill=black]
\tikzstyle{small white dot}=[small dot,fill=white]
\tikzstyle{small gray dot}=[small dot,fill=gray!40!white]

\tikzstyle{causal dot}=[inner sep=0.4mm,minimum width=0pt,minimum height=0pt,draw=white,shape=circle,fill=gray!40!white]

\tikzstyle{phase dimensions}=[minimum size=5mm,font=\footnotesize,rectangle,rounded corners=2.5mm,inner sep=0.2mm,outer sep=-2mm,text height=1ex, text depth=0.25ex, yshift=0.5mm]
\tikzstyle{dphase dimensions}=[phase dimensions]
\tikzstyle{phase dot}=[dot,phase dimensions]

\tikzstyle{white phase dot}=[dot,fill=white,phase dimensions]
\tikzstyle{white phase ddot}=[ddot,fill=white,dphase dimensions]

\tikzstyle{white rect ddot}=[draw=black,fill=white,doubled,minimum size=5mm,font=\footnotesize,rectangle,rounded corners=2.5mm,inner sep=0.2mm]
\tikzstyle{gray rect ddot}=[draw=black,fill=gray!40!white,doubled,minimum size=6mm,font=\footnotesize,rectangle,rounded corners=3mm]

\tikzstyle{gray phase dot}=[dot,fill=gray!40!white,phase dimensions]
\tikzstyle{gray phase ddot}=[ddot,fill=gray!40!white,dphase dimensions]
\tikzstyle{grey phase dot}=[gray phase dot]
\tikzstyle{grey phase ddot}=[gray phase ddot]

\tikzstyle{cnot}=[fill=white,shape=circle,inner sep=-1.4pt]
\tikzstyle{hadamard}=[square box,inner sep=0 pt,font=\footnotesize,minimum height=4mm,minimum width=4mm]
\tikzstyle{dhadamard}=[hadamard,doubled]
\tikzstyle{antipode}=[white dot,inner sep=0.3mm,font=\footnotesize]

\tikzstyle{scalar}=[diamond,draw,inner sep=0.5pt,font=\small]
\tikzstyle{dscalar}=[diamond,doubled, draw,inner sep=0.5pt,font=\small]

\tikzstyle{small box}=[rectangle,inline text,fill=white,draw,minimum height=5mm,yshift=-0.5mm,minimum width=5mm,font=\small]
\tikzstyle{small gray box}=[small box,fill=gray!30]
\tikzstyle{medium box}=[rectangle,inline text,fill=white,draw,minimum height=5mm,yshift=-0.5mm,minimum width=10mm,font=\small]
\tikzstyle{square box}=[small box] % for backwards-compatibility
\tikzstyle{medium gray box}=[small box,fill=gray!30]
\tikzstyle{semilarge box}=[rectangle,inline text,fill=white,draw,minimum height=5mm,yshift=-0.5mm,minimum width=12.5mm,font=\small]
\tikzstyle{large box}=[rectangle,inline text,fill=white,draw,minimum height=5mm,yshift=-0.5mm,minimum width=15mm,font=\small]
\tikzstyle{large gray box}=[small box,fill=gray!30]

\tikzstyle{gray square point}=[small box,fill=gray!50]

\tikzstyle{dphase box white}=[dbox]
\tikzstyle{dphase box gray}=[dbox,fill=gray!50!white]

\tikzstyle{point}=[regular polygon,regular polygon sides=3,draw,scale=0.75,inner sep=-0.5pt,minimum width=9mm,fill=white,regular polygon rotate=180]
\tikzstyle{copoint}=[regular polygon,regular polygon sides=3,draw,scale=0.75,inner sep=-0.5pt,minimum width=9mm,fill=white]
\tikzstyle{dpoint}=[point,doubled]
\tikzstyle{dcopoint}=[copoint,doubled]

\tikzstyle{wide copoint}=[fill=white,draw,shape=isosceles triangle,shape border rotate=90,isosceles triangle stretches=true,inner sep=0pt,minimum width=1.5cm,minimum height=6.12mm]
\tikzstyle{wide point}=[fill=white,draw,shape=isosceles triangle,shape border rotate=-90,isosceles triangle stretches=true,inner sep=0pt,minimum width=1.5cm,minimum height=6.12mm,yshift=-0.0mm]
\tikzstyle{wide point plus}=[fill=white,draw,shape=isosceles triangle,shape border rotate=-90,isosceles triangle stretches=true,inner sep=0pt,minimum width=1.74cm,minimum height=7mm,yshift=-0.0mm]

\tikzstyle{wide dpoint}=[fill=white,doubled,draw,shape=isosceles triangle,shape border rotate=-90,isosceles triangle stretches=true,inner sep=0pt,minimum width=1.5cm,minimum height=6.12mm,yshift=-0.0mm]

\tikzstyle{tinypoint}=[regular polygon,regular polygon sides=3,draw,scale=0.55,inner sep=-0.15pt,minimum width=6mm,fill=white,regular polygon rotate=180] 

\tikzstyle{white point}=[point]
\tikzstyle{white dpoint}=[dpoint]
\tikzstyle{green point}=[white point] % for backwards-compatibility
\tikzstyle{white copoint}=[copoint]
\tikzstyle{gray point}=[point,fill=gray!40!white]
\tikzstyle{gray dpoint}=[gray point,doubled]
\tikzstyle{red point}=[gray point] % for backwards-compatibility
\tikzstyle{gray copoint}=[copoint,fill=gray!40!white]
\tikzstyle{gray dcopoint}=[gray copoint,doubled]

\tikzstyle{black point}=[point,fill=black]
\tikzstyle{black copoint}=[copoint,fill=black]

\tikzstyle{tiny gray point}=[tinypoint,fill=gray!40!white]

\tikzstyle{diredge}=[->]
\tikzstyle{rdiredge}=[<-]
\tikzstyle{thickdiredge}=[->, very thick]
\tikzstyle{pointer edge}=[->,very thick,gray]
\tikzstyle{pointer edge part}=[very thick,gray]
\tikzstyle{dashed edge}=[dashed]
\tikzstyle{thick dashed edge}=[very thick,dashed]
\tikzstyle{thick gray dashed edge}=[thick dashed edge,gray!40]
\tikzstyle{thick map edge}=[very thick,|->]

\makeatletter
\newcommand{\boxshape}[3]{%
\pgfdeclareshape{#1}{
\inheritsavedanchors[from=rectangle] % this is nearly a rectangle
\inheritanchorborder[from=rectangle]
\inheritanchor[from=rectangle]{center}
\inheritanchor[from=rectangle]{north}
\inheritanchor[from=rectangle]{south}
\inheritanchor[from=rectangle]{west}
\inheritanchor[from=rectangle]{east}
\backgroundpath{% this is new
\southwest \pgf@xa=\pgf@x \pgf@ya=\pgf@y
\northeast \pgf@xb=\pgf@x \pgf@yb=\pgf@y

\@tempdima=#2
\@tempdimb=#3

\pgfpathmoveto{\pgfpoint{\pgf@xa - 5pt + \@tempdima}{\pgf@ya}}
\pgfpathlineto{\pgfpoint{\pgf@xa - 5pt - \@tempdima}{\pgf@yb}}
\pgfpathlineto{\pgfpoint{\pgf@xb + 5pt + \@tempdimb}{\pgf@yb}}
\pgfpathlineto{\pgfpoint{\pgf@xb + 5pt - \@tempdimb}{\pgf@ya}}
\pgfpathlineto{\pgfpoint{\pgf@xa - 5pt + \@tempdima}{\pgf@ya}}
\pgfpathclose
}
}}

\boxshape{NEbox}{0pt}{5pt}
\boxshape{SEbox}{0pt}{-5pt}
\boxshape{NWbox}{5pt}{0pt}
\boxshape{SWbox}{-5pt}{0pt}
\boxshape{EBox}{-3pt}{3pt}
\boxshape{WBox}{3pt}{-3pt}
\makeatother

\tikzstyle{cloud}=[shape=cloud,draw,minimum width=1.5cm,minimum height=1.5cm]

\tikzstyle{map}=[draw,shape=NEbox,inner sep=2pt,minimum height=6mm,fill=white]
\tikzstyle{dashedmap}=[draw,dashed,shape=NEbox,inner sep=2pt,minimum height=6mm,fill=white]
\tikzstyle{mapdag}=[draw,shape=SEbox,inner sep=2pt,minimum height=6mm,fill=white]
\tikzstyle{mapadj}=[draw,shape=SEbox,inner sep=2pt,minimum height=6mm,fill=white]
\tikzstyle{maptrans}=[draw,shape=SWbox,inner sep=2pt,minimum height=6mm,fill=white]
\tikzstyle{mapconj}=[draw,shape=NWbox,inner sep=2pt,minimum height=6mm,fill=white]

\tikzstyle{medium map}=[draw,shape=NEbox,inner sep=2pt,minimum height=6mm,fill=white,minimum width=7mm]
\tikzstyle{medium map dag}=[draw,shape=SEbox,inner sep=2pt,minimum height=6mm,fill=white,minimum width=7mm]
\tikzstyle{medium map adj}=[draw,shape=SEbox,inner sep=2pt,minimum height=6mm,fill=white,minimum width=7mm]
\tikzstyle{medium map trans}=[draw,shape=SWbox,inner sep=2pt,minimum height=6mm,fill=white,minimum width=7mm]
\tikzstyle{medium map conj}=[draw,shape=NWbox,inner sep=2pt,minimum height=6mm,fill=white,minimum width=7mm]
\tikzstyle{semilarge map}=[draw,shape=NEbox,inner sep=2pt,minimum height=6mm,fill=white,minimum width=9.5mm]
\tikzstyle{semilarge map trans}=[draw,shape=SWbox,inner sep=2pt,minimum height=6mm,fill=white,minimum width=9.5mm]
\tikzstyle{semilarge map adj}=[draw,shape=SEbox,inner sep=2pt,minimum height=6mm,fill=white,minimum width=9.5mm]
\tikzstyle{semilarge map dag}=[draw,shape=SEbox,inner sep=2pt,minimum height=6mm,fill=white,minimum width=9.5mm]
\tikzstyle{semilarge map conj}=[draw,shape=NWbox,inner sep=2pt,minimum height=6mm,fill=white,minimum width=9.5mm]
\tikzstyle{large map}=[draw,shape=NEbox,inner sep=2pt,minimum height=6mm,fill=white,minimum width=12mm]
\tikzstyle{very large map}=[draw,shape=NEbox,inner sep=2pt,minimum height=6mm,fill=white,minimum width=17mm]

\tikzstyle{medium dmap}=[draw,doubled,shape=NEbox,inner sep=2pt,minimum height=6mm,fill=white,minimum width=7mm]
\tikzstyle{medium dmap dag}=[draw,doubled,shape=SEbox,inner sep=2pt,minimum height=6mm,fill=white,minimum width=7mm]
\tikzstyle{medium dmap adj}=[draw,doubled,shape=SEbox,inner sep=2pt,minimum height=6mm,fill=white,minimum width=7mm]
\tikzstyle{medium dmap trans}=[draw,doubled,shape=SWbox,inner sep=2pt,minimum height=6mm,fill=white,minimum width=7mm]
\tikzstyle{medium dmap conj}=[draw,doubled,shape=NWbox,inner sep=2pt,minimum height=6mm,fill=white,minimum width=7mm]
\tikzstyle{semilarge dmap}=[draw,doubled,shape=NEbox,inner sep=2pt,minimum height=6mm,fill=white,minimum width=9.5mm]
\tikzstyle{semilarge dmap trans}=[draw,doubled,shape=SWbox,inner sep=2pt,minimum height=6mm,fill=white,minimum width=9.5mm]
\tikzstyle{semilarge dmap adj}=[draw,doubled,shape=SEbox,inner sep=2pt,minimum height=6mm,fill=white,minimum width=9.5mm]
\tikzstyle{semilarge dmap dag}=[draw,doubled,shape=SEbox,inner sep=2pt,minimum height=6mm,fill=white,minimum width=9.5mm]
\tikzstyle{semilarge dmap conj}=[draw,doubled,shape=NWbox,inner sep=2pt,minimum height=6mm,fill=white,minimum width=9.5mm]
\tikzstyle{large dmap}=[draw,doubled,shape=NEbox,inner sep=2pt,minimum height=6mm,fill=white,minimum width=12mm]
\tikzstyle{large dmap conj}=[draw,doubled,shape=NWbox,inner sep=2pt,minimum height=6mm,fill=white,minimum width=12mm]
\tikzstyle{large dmap trans}=[draw,doubled,shape=SWbox,inner sep=2pt,minimum height=6mm,fill=white,minimum width=12mm]
\tikzstyle{very large dmap}=[draw,doubled,shape=NEbox,inner sep=2pt,minimum height=6mm,fill=white,minimum width=19.5mm]

\tikzstyle{muxbox}=[draw,shape=rectangle,minimum height=3mm,minimum width=3mm,fill=white]
\tikzstyle{dmuxbox}=[muxbox,doubled]

\tikzstyle{box}=[draw,shape=rectangle,inner sep=2pt,minimum height=6mm,minimum width=6mm,fill=white]
\tikzstyle{dbox}=[draw,doubled,shape=rectangle,inner sep=2pt,minimum height=6mm,minimum width=6mm,fill=white]
\tikzstyle{dmap}=[draw,doubled,shape=NEbox,inner sep=2pt,minimum height=6mm,fill=white]
\tikzstyle{dmapdag}=[draw,doubled,shape=SEbox,inner sep=2pt,minimum height=6mm,fill=white]
\tikzstyle{dmapadj}=[draw,doubled,shape=SEbox,inner sep=2pt,minimum height=6mm,fill=white]
\tikzstyle{dmaptrans}=[draw,doubled,shape=SWbox,inner sep=2pt,minimum height=6mm,fill=white]
\tikzstyle{dmapconj}=[draw,doubled,shape=NWbox,inner sep=2pt,minimum height=6mm,fill=white]

\tikzstyle{ddmap}=[draw,doubled,dashed,shape=NEbox,inner sep=2pt,minimum height=6mm,fill=white]
\tikzstyle{ddmapdag}=[draw,doubled,dashed,shape=SEbox,inner sep=2pt,minimum height=6mm,fill=white]
\tikzstyle{ddmapadj}=[draw,doubled,dashed,shape=SEbox,inner sep=2pt,minimum height=6mm,fill=white]
\tikzstyle{ddmaptrans}=[draw,doubled,dashed,shape=SWbox,inner sep=2pt,minimum height=6mm,fill=white]
\tikzstyle{ddmapconj}=[draw,doubled,dashed,shape=NWbox,inner sep=2pt,minimum height=6mm,fill=white]

\boxshape{sNEbox}{0pt}{3pt}
\boxshape{sSEbox}{0pt}{-3pt}
\boxshape{sNWbox}{3pt}{0pt}
\boxshape{sSWbox}{-3pt}{0pt}
\tikzstyle{smap}=[draw,shape=sNEbox,fill=white]
\tikzstyle{smapdag}=[draw,shape=sSEbox,fill=white]
\tikzstyle{smapadj}=[draw,shape=sSEbox,fill=white]
\tikzstyle{smaptrans}=[draw,shape=sSWbox,fill=white]
\tikzstyle{smapconj}=[draw,shape=sNWbox,fill=white]

\tikzstyle{dsmap}=[draw,dashed,shape=sNEbox,fill=white]
\tikzstyle{dsmapdag}=[draw,dashed,shape=sSEbox,fill=white]
\tikzstyle{dsmaptrans}=[draw,dashed,shape=sSWbox,fill=white]
\tikzstyle{dsmapconj}=[draw,dashed,shape=sNWbox,fill=white]

\boxshape{mNEbox}{0pt}{10pt}
\boxshape{mSEbox}{0pt}{-10pt}
\boxshape{mNWbox}{10pt}{0pt}
\boxshape{mSWbox}{-10pt}{0pt}
\tikzstyle{mmap}=[draw,shape=mNEbox]
\tikzstyle{mmapdag}=[draw,shape=mSEbox]
\tikzstyle{mmaptrans}=[draw,shape=mSWbox]
\tikzstyle{mmapconj}=[draw,shape=mNWbox]

\tikzstyle{mmapgray}=[draw,fill=gray!40!white,shape=mNEbox]
\tikzstyle{smapgray}=[draw,fill=gray!40!white,shape=sNEbox]

\makeatletter
\pgfdeclareshape{cornerpoint}{
\inheritsavedanchors[from=rectangle] % this is nearly a rectangle
\inheritanchorborder[from=rectangle]
\inheritanchor[from=rectangle]{center}
\inheritanchor[from=rectangle]{north}
\inheritanchor[from=rectangle]{south}
\inheritanchor[from=rectangle]{west}
\inheritanchor[from=rectangle]{east}
\backgroundpath{% this is new
\southwest \pgf@xa=\pgf@x \pgf@ya=\pgf@y
\northeast \pgf@xb=\pgf@x \pgf@yb=\pgf@y

\pgfmathsetmacro{\pgf@shorten@left}{\pgfkeysvalueof{/tikz/shorten left}}
\pgfmathsetmacro{\pgf@shorten@right}{\pgfkeysvalueof{/tikz/shorten right}}

\pgfpathmoveto{\pgfpoint{0.5 * (\pgf@xa + \pgf@xb)}{\pgf@ya - 5pt}}
\pgfpathlineto{\pgfpoint{\pgf@xa - 8pt + \pgf@shorten@left}{\pgf@yb - 1.5 * \pgf@shorten@left}}
\pgfpathlineto{\pgfpoint{\pgf@xa - 8pt + \pgf@shorten@left}{\pgf@yb}}
\pgfpathlineto{\pgfpoint{\pgf@xb + 8pt - \pgf@shorten@right}{\pgf@yb}}
\pgfpathlineto{\pgfpoint{\pgf@xb + 8pt - \pgf@shorten@right}{\pgf@yb - 1.5 * \pgf@shorten@right}}
\pgfpathclose
}
}

\pgfdeclareshape{cornercopoint}{
\inheritsavedanchors[from=rectangle] % this is nearly a rectangle
\inheritanchorborder[from=rectangle]
\inheritanchor[from=rectangle]{center}
\inheritanchor[from=rectangle]{north}
\inheritanchor[from=rectangle]{south}
\inheritanchor[from=rectangle]{west}
\inheritanchor[from=rectangle]{east}
\backgroundpath{% this is new
\southwest \pgf@xa=\pgf@x \pgf@ya=\pgf@y
\northeast \pgf@xb=\pgf@x \pgf@yb=\pgf@y

\pgfmathsetmacro{\pgf@shorten@left}{\pgfkeysvalueof{/tikz/shorten left}}
\pgfmathsetmacro{\pgf@shorten@right}{\pgfkeysvalueof{/tikz/shorten right}}

\pgfpathmoveto{\pgfpoint{0.5 * (\pgf@xa + \pgf@xb)}{\pgf@yb + 5pt}}
\pgfpathlineto{\pgfpoint{\pgf@xa - 8pt + \pgf@shorten@left}{\pgf@ya + 1.5 * \pgf@shorten@left}}
\pgfpathlineto{\pgfpoint{\pgf@xa - 8pt + \pgf@shorten@left}{\pgf@ya}}
\pgfpathlineto{\pgfpoint{\pgf@xb + 8pt - \pgf@shorten@right}{\pgf@ya}}
\pgfpathlineto{\pgfpoint{\pgf@xb + 8pt - \pgf@shorten@right}{\pgf@ya + 1.5 * \pgf@shorten@right}}
\pgfpathclose
}
}

\makeatother

\pgfkeyssetvalue{/tikz/shorten left}{0pt}
\pgfkeyssetvalue{/tikz/shorten right}{0pt}

\tikzstyle{kpoint common}=[draw,fill=white,inner sep=1pt,minimum height=3mm]
\tikzstyle{kpoint}=[shape=cornerpoint,shorten left=5pt,kpoint common]
\tikzstyle{kpoint adjoint}=[shape=cornercopoint,shorten left=5pt,kpoint common]
\tikzstyle{kpoint conjugate}=[shape=cornerpoint,shorten right=5pt,kpoint common]
\tikzstyle{kpoint transpose}=[shape=cornercopoint,shorten right=5pt,kpoint common]
\tikzstyle{kpoint symm}=[shape=cornerpoint,shorten left=5pt,shorten right=5pt,kpoint common]

\tikzstyle{black kpoint}=[shape=cornerpoint,shorten left=5pt,kpoint common,fill=black]
\tikzstyle{black kpoint adjoint}=[shape=cornercopoint,shorten left=5pt,kpoint common,fill=black]

\tikzstyle{kpointdag}=[kpoint adjoint]
\tikzstyle{kpointadj}=[kpoint adjoint]
\tikzstyle{kpointconj}=[kpoint conjugate]
\tikzstyle{kpointtrans}=[kpoint transpose]

\tikzstyle{big kpoint}=[kpoint, minimum width=1.2 cm, minimum height=8mm, inner sep=4pt, text depth=3mm]

\tikzstyle{wide kpoint}=[kpoint, minimum width=1 cm, inner sep=2pt, text depth=-0.7 mm]
\tikzstyle{wide kpointdag}=[kpointdag, minimum width=1 cm, inner sep=2pt, text depth=0.7 mm]
\tikzstyle{wide kpointconj}=[kpointconj, minimum width=1 cm, inner sep=2pt, text depth=-0.7 mm]
\tikzstyle{wide kpointtrans}=[kpointtrans, minimum width=1 cm, inner sep=2pt, text depth=0.7 mm]

\tikzstyle{gray kpoint}=[kpoint,fill=gray!50!white]
\tikzstyle{gray kpointdag}=[kpointdag,fill=gray!50!white]
\tikzstyle{gray kpointadj}=[kpointadj,fill=gray!50!white]
\tikzstyle{gray kpointconj}=[kpointconj,fill=gray!50!white]
\tikzstyle{gray kpointtrans}=[kpointtrans,fill=gray!50!white]

\tikzstyle{gray dkpoint}=[kpoint,fill=gray!50!white,doubled]
\tikzstyle{gray dkpointdag}=[kpointdag,fill=gray!50!white,doubled]
\tikzstyle{gray dkpointadj}=[kpointadj,fill=gray!50!white,doubled]
\tikzstyle{gray dkpointconj}=[kpointconj,fill=gray!50!white,doubled]
\tikzstyle{gray dkpointtrans}=[kpointtrans,fill=gray!50!white,doubled]

\tikzstyle{white label}=[draw,fill=white,rectangle,inner sep=0.7 mm]
\tikzstyle{gray label}=[draw,fill=gray!50!white,rectangle,inner sep=0.7 mm]
\tikzstyle{black label}=[draw,fill=black,rectangle,inner sep=0.7 mm]

\tikzstyle{dkpoint}=[kpoint,doubled]
\tikzstyle{wide dkpoint}=[wide kpoint,doubled]
\tikzstyle{dkpointdag}=[kpoint adjoint,doubled]
\tikzstyle{dkcopoint}=[kpoint adjoint,doubled]
\tikzstyle{dkpointadj}=[kpoint adjoint,doubled]
\tikzstyle{dkpointconj}=[kpoint conjugate,doubled]
\tikzstyle{dkpointtrans}=[kpoint transpose,doubled]

\tikzstyle{kscalar}=[kpoint common, shape=EBox, inner xsep=-1pt, inner ysep=3pt,font=\small]
\tikzstyle{kscalarconj}=[kpoint common, shape=WBox, inner xsep=-1pt, inner ysep=3pt,font=\small]

 \tikzstyle{upground}=[circuit ee IEC,thick,ground,rotate=90,scale=2.5]
 \tikzstyle{downground}=[circuit ee IEC,thick,ground,rotate=-90,scale=2.5]
 \tikzstyle{bigground}=[regular polygon,regular polygon sides=3,draw=gray,scale=0.50,inner sep=-0.5pt,minimum width=10mm,fill=gray]
\tikzstyle{arrs}=[-latex,font=\small,auto]
\tikzstyle{arrow plain}=[arrs]
\tikzstyle{arrow dashed}=[dashed,arrs]
\tikzstyle{arrow bold}=[very thick,arrs]
\tikzstyle{arrow hide}=[draw=white!0,-]
\tikzstyle{arrow reverse}=[latex-]
\tikzstyle{cdnode}=[]

\newcommand{\stdpart}[1]{\operatorname{st}(#1)}
\newcommand{\indexSet}[1]{\{1,...,\dim{#1}\}}
\newcommand{\truncate}[1]{\bar{#1}}
\newcommand{\starHilbCategoryNearStd}{\starHilbCategory^{(std)}}
\newcommand{\vartruncate}[1]{\operatorname{trunc}[#1]}

\title{Infinite-dimensional Categorical Quantum Mechanics}
\author{
	Stefano Gogioso
	\institute{Quantum Group \\ University of Oxford}
	\email{stefano.gogioso@cs.ox.ac.uk}
	\and
	Fabrizio Genovese
	\institute{Quantum Group \\ University of Oxford}
	\email{fabrizio.genovese@hertford.ox.ac.uk}
}
\def\titlerunning{Infinite-dimensional Categorical Quantum Mechanics}
\def\authorrunning{S. Gogioso \& F. Genovese}

\begin{document}

\bibliographystyle{eptcs}

\maketitle

\begin{abstract}
We use non-standard analysis to define a category $\starHilbCategory$ suitable for categorical quantum mechanics in arbitrary separable Hilbert spaces, and we show that standard bounded operators can be suitably embedded in it. We show the existence of unital special commutative $\dagger$-Frobenius algebras, and we conclude $\starHilbCategory$ to be compact closed, with partial traces and a Hilbert-Schmidt inner product on morphisms. We exemplify our techniques on the textbook case of 1-dimensional wavefunctions with periodic boundary conditions: we show the momentum and position observables to be well defined, and to give rise to a strongly complementary pair of unital commutative $\dagger$-Frobenius algebras. 
\end{abstract}

\section{Introduction} 

%% Issues with infinite-dimensional CQM 
Throughout the past decade, the framework of categorical quantum mechanics (CQM)\cite{abramsky2009categorical,coecke2011interacting,coecke2008classical} has achieved remarkable success in describing the foundations of finite-dimensional quantum theory, and the structures behind quantum information protocols and quantum computation. Unfortunately, attempts to extend the same techniques to the treatment of infinite-dimensional case have so far achieved limited success. Although the work of \cite{abramsky2012hstaralgebras} on H$^\star$-algebras provides a characterisation of non-degenerate observables in arbitrary dimensions, much of the traditional machinery of CQM is nonetheless lost, e.g.: 
\vspace{-0.5mm}
\begin{enumerate}
\itemsep-0.2em
\item[(a)] dagger compact structure: operator-state duality and Hilbert-Schmidt inner product;
\item[(b)] Hopf algebras and complementarity: mutually unbiased observables;
\item[(c)] strong complementarity: quantum symmetries and the position/momentum observables.
\end{enumerate}
\vspace{-0.5mm}

%% NSA to deal with unbounded operators
\noindent In this work, we resort to non-standard analysis \`{a} la Robinson \cite{robinson1974nonstandard} to tackle the issue of infinitesimal and infinite quantities behind unbounded operators, Dirac deltas and plane-waves: these are key ingredients of mainstream quantum mechanics which the categorical framework has thus failed to adequately capture, and we demonstrate how they can be used to recover a great deal of CQM machinery in infinite-dimensions. Applications of non-standard analysis to quantum theory already appeared in the past decades \cite{ozawa1993unitary,Farrukh1975application}, but in a different spirit and with different objectives in mind.

%% Synopsis
In Section~\ref{section_NonStandardAnalysis}, we provide a basic summary of the non-standard techniques we will be using. In Section~\ref{section_StarHilbCategory}, we construct a category $\starHilbCategory$ of non-standard separable Hilbert spaces, and we relate it to the category $\sHilbCategory$ of standard separable Hilbert spaces and bounded linear maps. In Section~\ref{section_InfiniteDimCQM} we use our newly defined category to extend CQM from finite to separable Hilbert spaces, and we treat the textbook case of position and momentum observable for 1-dimensional wavefunctions with periodic boundary conditions. 

%% CQM for 1-dimensional periodic wavefunctions
We begin by showing the existence of unital special commutative $\dagger$-Frobenius algebras, and we deduce that $\starHilbCategory$ is a dagger compact category. From a countable basis of momentum eigenstates, we define the position eigenstates as Dirac deltas, and construct the position and momentum observables as a unital special commutative $\dagger$-Frobenius algebras. Furthermore, we show these observables to be strongly complementary: this is the categorical counterpart of the Weyl canonical commutation relations, and opens the way to future applications of the formalism to infinite-dimensional quantum symmetries and dynamics (within the framework of \cite{gogioso2015categorical}).

\section{Non-standard analysis}
\label{section_NonStandardAnalysis}

\vspace{-0.1mm}

\subsection{Non-standard models}

%% Higher order logic
Non-standard analysis has its roots in the field of higher order logic \cite{robinson1974nonstandard}, where universal and existential quantifiers are not only allowed to range over individuals, but also over relations between individuals, relations between relations, and so on. A model of a higher order theory can then be thought as a family $M=\{B_\tau\}_{\tau \in T}$ of sets, indexed by a set $T$ of types, with $B_\tau$ containing all relations of type $\tau$ in the model.

%% Non-standard models and transfer theorem
Given a higher order theory, with a (standard) model $M=\{B_\tau\}_{\tau \in T}$, the ultraproduct construction from \cite{robinson1974nonstandard} produces a \textbf{non-standard model}\footnote{Non-standard models are denoted by a prefix $^\star$, bearing no relation to the postfix $^\star$ of complex conjugation.} $\nonstd{M} = \{\nonstd{B_\tau}\}_{\tau \in T}$. The set $\nonstd{B_\tau}$ of relations of type $\tau$ in the non-standard model contains $B_\tau$ as a subset, and in particular the individuals $B_0$ of the standard model form a subset of the individuals $\nonstd{B_0}$ of the non-standard model. The non-standard model, however, is not \textit{full}: the set $\nonstd{B_\tau}$ does not contain, in general, all set-theoretic relations of type $\tau$ that can be constructed from the non-standard individuals. In order to distinguish the relations in $\nonstd{B_\tau}$ from the larger family of set theoretic ones, we will refer to the former as \textbf{internal} relations, and to the latter as \textbf{external} relations. Also, we will refer to the relations in $B_\tau$ as \textbf{standard} relations, and we will freely confuse them with the ones in the standard model.
The main result used to prove existence and properties of internal individuals/relations in a non-standard model is called \textbf{transfer theorem}, and can be summarised as follows. 
\begin{theorem}[Transfer Theorem]
Let $\Phi$ be a higher order formula which is admissible\footnote{I.e. one which does not contain any non-standard individual/relational symbols.} in the higher order theory. Then $\Phi$ holds in the standard model $M$ with quantifiers ranging over standard relations if and only if $\Phi$ holds in the non-standard model $\nonstd{M}$ with quantifiers ranging over internal relations.
\end{theorem}
\noindent In the forward direction, the transfer theorem says that every property of standard individuals/relations in the standard model applies to internal individuals/relations in the non-standard model. In the backward direction (which will not play a direct role in this work), the transfer theorem says that any statement holding in $\nonstd{M}$ also holds in $M$, as long as it is \textit{admissible} in $M$.

\vspace{-0.1mm}

\subsection{The structure of \texorpdfstring{$\starNaturals$}{*N}}

%% Finite and infinite naturals
The \textbf{non-standard naturals} $\starNaturals$ form a totally ordered semiring, with the \textbf{standard naturals} $\naturals$ as an initial segment. As a totally ordered set, the non-standard naturals are order-isomorphic to $\naturals + \theta \times \integers$, where $\theta$ a dense order with no maximum nor minimum. We refer to the standard naturals as \textbf{finite naturals}, and to the internal naturals in $\starNaturals - \naturals$ as \textbf{infinite naturals}: this is because any infinite natural $\kappa$ satisfies $\kappa > n$ for all $n \in \naturals$. We say that two non-standard naturals $n,m$ have the same \textbf{order of infinity} if they differ by a finite natural $|n-m| \in \naturals$: this gives an equivalence relation, and the set of equivalence classes is in order-preserving bijection with the totally ordered set $\Theta^+ := \{0\} + \theta$. The set $\Theta^+$ also inherits the additive monoid structure of $\starNaturals$, but not the full semiring structure.

%% Induction
By transfer theorem, many properties of $\naturals$ transfer to $\starNaturals$: for example, from the fact that every non-empty set of standard naturals has a minimum we conclude that every non-empty internal set of non-standard naturals also has a minimum, and arguments by induction can be carried out on non-empty internal subsets of $\starNaturals$. However, the requirement that the set be internal is key: the set of all infinite naturals, for example, has no minimum (and hence it cannot be internal). 

%% Sequences of naturals
If $(a_n)_{n\in\naturals}$ is a sequence of natural numbers defined by some formula $\forall n \in \naturals \, \exists \, a_n \in \naturals \, \Phi(n,a_n)$ in the standard model, then by transfer theorem there exists a unique corresponding standard sequence $(a_n)_{n \in \starNaturals}$ in the non-standard model, coinciding with the $(a_n)_{n \in \naturals}$ for all finite naturals. Furthermore, for any $m \in \naturals$ the naturals $s_m := \sum_{n=0}^m a_n$ and $p_m := \prod_{n=0}^m a_n$ exist in the standard model, and hence the non-standard naturals $s_m$ and $p_m$ exist in the non-standard model for all $m \in \starNaturals$: for example, if $(a_n)_{n \in \naturals}$ is the sequence of finite primes, then $p_\kappa := \prod_{n=0}^\kappa a_n$ for any infinite natural $\kappa$ exists, divisible by all finite primes.

%% Non-standard integers
The \textbf{non-standard integers} $\starIntegers$ similarly relate to the standard integers $\integers$: they form a totally ordered ring, with $\integers$ as a sub-ring and $\starNaturals$ as a sub-semiring. As a totally ordered set, they are order-isomorphic to $(\theta + \{0\} + \theta) \times \integers$: they contain the finite integers together two copies of the infinite naturals, one copy above all finite integers (the \textbf{positive infinities}) and one copy below all finite integers (the \textbf{negative infinities}). The set $\Theta := \theta + \{0\} + \theta$ of orders of infinity for $\starIntegers$ again inherits the total order and the additive group structure, but not the ring one. 

\subsection{The structure of \texorpdfstring{$\starReals$}{*R}}
\label{section_StructureStarReals}

%% Finite, infinite and infinitesimal reals
The \textbf{non-standard reals} $\starReals$ form an ordered field, with the \textbf{standard reals} $\reals$ as a sub-field and the non-standard integers $\starIntegers$ as a subring. They are a non-archimedean field, with a sub-ring $M_1$ of \textbf{infinitesimals}, smaller in absolute value than all positive standard reals. The non-zero infinitesimals have inverses, the \textbf{infinite reals}, larger in absolute value than all positive standard integers/reals. 

%% Standard part
By using the finite integers $\integers \subset \starIntegers \subset \starReals$, it is possible to define the sub-ring\footnote{In fact, they form a $\reals$-vector subspace of $\starReals$.} $M_0$ of the \textbf{finite reals}, given by those $x \in \starReals$ such that $\exists \; n \in \integers \; |x| < n$. The sub-ring $M_1$ of infinitesimals is a two-sided ideal in $M_0$, and by using Dedekind cuts it is possible to show that the ring quotient $M_0 / M_1$ is isomorphic to $\reals$: we refer to the corresponding surjective ring homomorphism $\stdpart{\emptyArg}: M_0 \rightarrow \reals$ as the \textbf{standard part} (which is the identity on the subring $\reals \leq M_0$), and we denote the corresponding quotient equivalence relation on $M_0$ by $x \simeq y \iffdef |x-y| \text{ is an infinitesimal}$. The coset of $M_1$ surrounding any non-standard real $x \in \starReals$, called the \textbf{monad} of $x$, and when $x$ is finite it contains exactly one standard real $\stdpart{x} \in \reals$.

% Density of standard rationals, and structure of standard reals
The non-standard reals are Archimedean in a non-standard sense: by transfer theorem, for any $x \in \starReals$ there is a unique $n \in \starNaturals$ such that $n \leq |x| < n+1$.\footnote{Equivalently, for every infinitesimal $\xi \in M_1$ there is a unique non-standard natural $n \in \starNaturals$ such that $1/(n+1) < |\xi| \leq 1/n$.} As a consequence, the non-standard rationals $\starRationals$ are a dense subfield of $\starReals$. Furthermore, the non-standard reals can be obtained in the familiar way by \inlineQuote{gluing} a copy of the (non-standard) unit interval between any two consecutive (non-standard) integers: as a totally ordered additive group, they are then isomorphic to $\Theta \times M_0$. 

%% Limits of sequences
Any sequence $(a_n)_{n \in \naturals}$ of reals definable in the standard model has a corresponding non-standard extension $(a_n)_{n \in \starNaturals}$ by transfer theorem: it coincides with the original sequence on all finite naturals, but will not in general be valued in the standard reals on infinite naturals. It is possible to show that $\lim_{n \rightarrow \infty} a_n = a \in \reals$ in the standard model if and only if $a_n \simeq a$ for all infinite naturals $n$ in the non-standard model. Furthermore, $(a_n)_{n \in \naturals}$ is bounded (say by $|a_n| \leq z \in \reals^+$) in the standard model if and only if $a_n$ is a finite real (with $|a_n| \leq z$) for all infinite naturals in the non-standard model.

%% Limits and continuity of functions
Real-valued functions $f: I \rightarrow \reals$ in the standard model can similarly be extended by transfer theorem to real-valued $f: \nonstd{I} \rightarrow \starReals$ in the non-standard model, coinciding with the original function on all standard reals in $\nonstd{I}$. Then $\lim_{x \rightarrow a} f(x) = c$ in the standard model if and only if in the non-standard model we have $f(x) \simeq c$ for all $x \simeq a$ (except perhaps at $x = a$). As a consequence, $f$ is continuous at $a \in I$ in the standard model if and only if in the non-standard model we have $f(x) \simeq f(a)$ whenever $x \simeq a$. 

%% Non-standard complex numbers
The \textbf{non-standard complex numbers} $\starComplexs$ similarly extend $\complexs$ with infinitesimals and infinities: they also form a field, with both $\starReals$ and $\complexs$ as sub-fields. As an additive group, they are isomorphic to $\starReals^2$. We will transfer most notations from $\starReals$ to $\starComplexs$, when no confusion can arise.

\subsection{Non-standard Hilbert spaces}

%% The effect of non-standard analysis on Hilbert spaces
The passage from standard to non-standard models has a two-fold effect on (complex) Hilbert spaces: (i) the scalars change from $\complexs$ to $\starComplexs$; (ii) the vectors change from sequences $(a_n)_{n \in \naturals^+}$ indexed by the standard naturals to sequences $(a_n)_{n \in \starNaturals^+}$ indexed by the non-standard naturals. Each standard Hilbert space $V$ has a non-standard counterpart $\nonstd{V}$: the non-standard space $\nonstd{V}$ containing all vectors of $V$, known as the \textbf{standard vectors}, as a $\complexs$-linear (but not $\starComplexs$-linear) subspace. The non-standard space $\nonstd{V}$ comes with a $\starComplexs$-valued inner product (extending the standard one on $V$), and an associated $\starReals^+$-valued norm. 

%% Infinitesimal vetors
The vectors infinitesimally close to standard vectors are called \textbf{near-standard vectors}, and the vectors with infinitesimal norm are called \textbf{infinitesimal vectors}: both form $\complexs$-linear (and $M_0$-linear) subspaces $\nonstd{V_0}$ and $\nonstd{V_1}$ of $\nonstd{V}$. There is a $\complexs$-linear map $\stdpart{\emptyArg}: \nonstd{V}_0 \rightarrow V$, known as the \textbf{standard part}, which sends the near-standard vectors surjectively onto $V$, acts as the identity on standard vectors and has the infinitesimal vectors $V_1$ as kernel. The standard part defines an equivalence relation $\simeq$ on near-standard vectors, with $\ket{\psi} \simeq \ket{\phi} $ if and only if $\ket{\psi} - \ket{\phi}$ is an infinitesimal vector. 

%% \kappa-truncated plain waves
An interesting class of non-standard vectors can be obtained by transfer theorem. Consider a standard complex Hilbert space $V$ which is separable, i.e. comes with a complete orthonormal basis $\ket{e_{n}}_{n \in \naturals^+}$ which is countable\footnote{We index our vectors in the positive naturals $\naturals^+$ for reasons of convenience: this way a generic vector in a $d$-dimensional vector space is written cleanly as $\sum_{n=1}^d v_n \ket{e_n}$.}. If $(a_{n})_{n \in \naturals^+}$ is a standard sequence of complex numbers, then the vector $\ket{\psi^{(k)}} := \sum_{n=1}^{k} a_n \ket{e_n} \in V$ exists for all positive standard naturals $k \in \naturals^+$: by transfer theorem, the vector $\ket{\psi^{(\kappa)}}$ exists in $\nonstd{V}$ for any infinite natural $\kappa$, where the corresponding non-standard sequence $(a_n)_{n \in \starNaturals^+}$ is used to provide values. In particular, the vector $\sum_{n=1}^{\kappa} \ket{e_n} \in \nonstd{V}$ exists, and has squared norm $\kappa \in \starReals^+$. 

The vectors of finite norm are known as \textbf{finite vectors} and form a $\complexs$-linear (and $M_0$-linear) subspace of $V$. However, this is where the second effect of non-standard analysis on Hilbert spaces comes into play: there exist finite vectors, such as $\ket{\phi} := \frac{1}{\sqrt{\kappa}} \sum_{n=1}^{\kappa} \ket{e_n}$, which are not near-standard. Indeed, any standard vector $\ket{\psi}$ is infinitesimally close to its truncation in the form $\ket{\psi^{(\kappa)}} := \sum_{n=1}^{\kappa} \psi_n \ket{e_n}$, where $\psi_\nu$ is infinitesimal for all infinite naturals $\nu$. We get the following lower bound for the squared norm of the difference $\ket{\phi}-\ket{\psi}$:
\begin{align}
\Big|\Big| \ket{\phi}-\ket{\psi} \Big|\Big|^2 \simeq \Big|\Big| \ket{\phi}-\ket{\psi^{(\kappa)}} \Big|\Big|^2 = &\sum_{n=1}^{\kappa} \frac{|1-\psi_n|}{\kappa}^2 \geq \sum_{n\geq\kappa/M} \frac{|1-\psi_n|^2}{\kappa} \nonumber \\
\geq &\sum_{n\geq\kappa/M} \frac{1-\epsilon}{\kappa} = (1-\epsilon)(1-\frac{1}{M}) \;\;\text{ for all $M \in \naturals^{+}$ and $\epsilon \in (0,1)$} \nonumber
\end{align}
This means that $\stdpart{\big|\big| \ket{\phi}-\ket{\psi^{(\kappa)}} \big|\big|} \geq 1$ for all standard vectors $\ket{\psi}$, and hence the vector $\ket{\phi}$ is finite but not near-standard. Finite vectors which are not near-standard are genuinely new, and can be used to do genuinely new things. This is what makes the non-standard approach to quantum mechanics so powerful: in $\starReals$ and $\starComplexs$ finite numbers are all near-standard, and correspond to standard numbers under infinitesimal equivalence, while in a non-standard Hilbert space one gets new things for free, such as normalised plane-waves in $\Ltwo{\integers}$ and Dirac-deltas in $\Ltwo{\reals/(L\integers)}$. These will be the fundamental building blocks of our work.

%% Transfer theorem to define internal morphisms
The transfer theorem can similarly be used to define non-standard linear operators (not necessarily continuous/bounded): if $(a_{nm})_{n, m \in \naturals^+}$ is a doubly-indexed sequence (a.k.a. a matrix) of complex numbers, then the linear operator $\sum_{m,n=0}^{\kappa} a_{mn} \, \ket{e_m}\bra{e_n} : \nonstd{V} \,\rightarrow \nonstd{V}$ exists for any infinite natural $\kappa$ (where $(a_{nm})_{n, m \in \starNaturals^+}$ is the unique internal non-standard sequence given by transfer theorem). This is a remarkable result, but it comes with some tricky limitations which will be presented in the next section.

\newpage
\section{The category \texorpdfstring{$\starHilbCategory$}{*Hilb}}
\label{section_StarHilbCategory}
The main idea behind our construction is to legitimise, through non-standard analysis, notations such as $\sum_{n\in \naturals^+} \ket{e_n}\bra{e_n}$ for the identity operator, $\sum_{n\in\naturals^+} \ket{e_n}$ for the unit of an infinite-dimensional Frobenius algebra, $\sum_{n,m \in \naturals^+}\ket{e_n}a_{nm}\bra{a_m}$ for a general matrix $(a_{nm})_{n,m \in \naturals^+}$. The transfer theorem doesn't allow us to conclude the existence of sums strictly over $\naturals^+$ (nor over the entirety of $\starNaturals$), but it does allow us to sum up to some infinite natural $\kappa$: the sums $\sum_{n=1}^\kappa \ket{e_n}\bra{e_n}$, $\sum_{n=1}^\kappa \ket{e_n}$ and $\sum_{n,m=1}^\kappa \ket{e_n} a_{nm} \bra{e_m}$ all describe well-defined internal linear maps of non-standard Hilbert spaces. Unfortunately,  $P_\kappa := \sum_{n=1}^\kappa \ket{e_n}\bra{e_n}$ does not behave like the identity over the space of all internal linear maps, but rather it is as a subspace projector: in order to turn these projectors into identities, we use a construction similar to that of the Cauchy/idempotent\footnote{Projectors are self-adjoint idempotents.} completion. As it turns out, this procedure preserves all standard bounded operators, and enough non-standard ones to do many of the things we care about in categorical quantum mechanics.

\subsection{Definition of the category}
\label{section_StarHilbCategorydef}

%% Objects
We proceed to define the \textbf{category of non-standard separable Hilbert spaces}\footnote{We have complex Hilbert spaces in mind, but the construction is identical for real Hilbert spaces.}, which we will denote by $\starHilbCategory$. All proofs of results in this and future sections can be found in the Appendix. As objects we take separable (standard) Hilbert spaces together with a witness of separability, i.e. pairs $\SpaceH := \big(V, \ket{e_{n}}_{n=1}^\kappa\big)$ of a standard separable Hilbert space $V$ and either
\begin{enumerate}
\item[(i)] if $V$ is finite-dimensional: a finite orthonormal basis $\ket{e_{n}}_{n=1}^\kappa$, where $\kappa := \dim{V} \in \naturals$;
\item[(ii)] if $V$ is infinite-dimensional: the unique extension (by transfer theorem) up to some infinite natural $\kappa \in \,\starNaturals$ of a complete countable orthonormal basis $\ket{e_{n}}_{n\in \naturals^+}$ for $V$. 
\end{enumerate} 

%% Truncating projectors and morphisms
\noindent For each object $\SpaceH := \big(V, \ket{e_{n}}_{n=1}^\kappa\big)$, let the \textbf{truncating projector} $P_\SpaceH: \SpaceH \rightarrow \SpaceH$ be the following internal linear map $\nonstd{V} \rightarrow \nonstd{V}$, where we refer to $\dim{\SpaceH} := \kappa \in \starNaturals$ as the \textbf{dimension} of object $\SpaceH$:
\begin{equation}\label{eqn_TruncatingProjector}
P_\SpaceH := \sum_{n=1}^{\dim{\SpaceH}} \ket{e_n} \bra{e_n}.
\end{equation}

\noindent We also use notation $|\SpaceH| := V$ to refer to the standard separable Hilbert space underlying an object $\SpaceH$ of $\starHilbCategory$. The morphisms in the category $\starHilbCategory$ are then defined as follows:
\begin{equation}\label{eqn_Homset}
\Hom{\starHilbCategory}{\SpaceH}{\SpaceG} := \suchthat{\;P_\SpaceG \circ F \circ P_\SpaceH\;}{\;F:\nonstd{|\SpaceH|} \,\rightarrow\, \nonstd{|\SpaceG|} \text{ internal linear map}}.
\end{equation}

\noindent Because the truncating projectors for $\SpaceH$ and $\SpaceG$ are internal linear maps, the composite $P_\SpaceG \circ F \circ P_\SpaceH$ is an internal linear map $\nonstd{|\SpaceH|} \, \rightarrow \nonstd{|\SpaceG|}$, which we shall denote by $\truncate{F}$. Composition of morphisms in $\starHilbCategory$ is simply composition of internal linear maps
\begin{equation}\label{eqn_Composition}
\truncate{G} \cdot \truncate{F} := \truncate{G} \circ \truncate{F} = (P_\SpaceG \circ G \circ P_\SpaceH) \circ (P_\SpaceH \circ F  \circ P_\SpaceK) = P_\SpaceG \circ (G \circ  P_\SpaceH \circ F)  \circ P_\SpaceK,
\end{equation}
where we used associativity of composition and idempotence of truncating projectors. Idempotence of the projectors, in particular, means that they provide suitable identity morphisms. Indeed if we define
\begin{equation}
    \id{\SpaceH} := P_\SpaceH \circ \id{\nonstd{|\SpaceH|}} \circ P_\SpaceH = P_\SpaceH \circ P_\SpaceH = P_\SpaceH,
\end{equation}
it is straightforward to check that $\id{\SpaceG} \cdot  \truncate{F} = P_\SpaceG \circ P_\SpaceG \circ F \circ P_\SpaceH = P_\SpaceG \circ F \circ P_\SpaceH = \truncate{F}$, and similarly for $\truncate{F} \cdot \id{\SpaceH}$.

\newpage
%% Tensor product
\noindent Now consider two naturals $\kappa, \nu \in \starNaturals$, and define the internal map
\begin{equation}\label{eqn_varsigma}
\varsigma_{\kappa,\nu}(n,m) := (n-1)\nu + m,
\end{equation}
which is an internal bijection between $\{1,...,\kappa\}\times\{1,...,\nu\}$ and $\{1,...,\kappa \nu\}$. Also, we will simply write $\varsigma(n,m)$ when no confusion can arise. A tensor product can be defined on the objects of $\starHilbCategory$ as follows, with tensor unit $(\complexs,1)$:
\begin{equation}\label{eqn_TensorObjects}
\Big(V, \ket{e_{n}}_{n=1}^\kappa\Big)
\otimes \Big(W, \ket{f_{m}}_{m=1}^\nu\Big) 
:= \Big(V \otimes W, \big(\ket{e_{n}} \otimes \ket{f_{m}}\big)_{\varsigma(n,m)=1}^{\kappa\nu}\Big).
\end{equation}

\noindent In order to define the tensor product on morphisms, we need to first note that morphisms $\truncate{F} : \SpaceH \rightarrow \SpaceG$ in $\starHilbCategory$ are uniquely determined by certain matrices $\indexSet{\SpaceG} \times \indexSet{\SpaceH} \rightarrow\, \starComplexs$:
\begin{equation}\label{eqn_MatrixRepresentation}
\truncate{F} = P_\SpaceG \circ F \circ P_\SpaceH = 
\sum_{m=1}^{\dim{\SpaceG}} \sum_{n=1}^{\dim{\SpaceH}} 
\ket{f_{m}} \Big( \bra{f_{m}} F \ket{e_{n}} \Big) \bra{e_{n}}.
\end{equation}

\noindent We introduce the notation $\truncate{F}_{mn} := \bra{f_m} F \ket{e_n}$, and define the tensor product of two morphisms $\truncate{F} : \SpaceH \rightarrow \SpaceG$ and $\truncate{G} : \SpaceH' \rightarrow \SpaceG'$ to be the familiar tensor product of matrices:
\begin{equation}\label{eqn_TensorMorphisms}
\truncate{F} \otimes \truncate{G} := 
\sum_{\varsigma(m,m')=1}^{\dim{\SpaceG}\dim{\SpaceG'}}
\sum_{\varsigma(n,n')=1}^{\dim{\SpaceH}\dim{\SpaceH'}}
\ket{f_{m}} \otimes \ket{f'_{m'}}  \; \truncate{F}_{mn} \truncate{G}_{m'n'} \; \bra{e_{n}} \otimes \bra{e'_{n'}}.
\end{equation}

\noindent The map $\truncate{F} \otimes \truncate{G}$ is an internal linear map $\nonstd{|\SpaceH|} \,\otimes\, \nonstd{|\SpaceH'|} \rightarrow \nonstd{|\SpaceG|} \,\otimes\, \nonstd{|\SpaceG'|}$ by transfer theorem. Also we have that $P_\SpaceH \otimes P_{\SpaceH'} = P_{\SpaceH \otimes \SpaceH'}$, and that $\truncate{F} \otimes \truncate{G} = P_{\SpaceG \otimes \SpaceG'} \circ \big(\truncate{F} \otimes \truncate{G}\big) \circ P_{\SpaceH \otimes \SpaceH'}$. Hence, $\truncate{F} \otimes \truncate{G}$ is a genuine morphism $\SpaceH \otimes \SpaceH' \rightarrow \SpaceG \otimes \SpaceG'$. It is straightforward to check that this results in a well defined tensor product\footnote{An elementary proof of associativity is provided in the appendix.}, and the following braiding operator turns $\starHilbCategory$ into a symmetric monoidal category (SMC):
\begin{equation}\label{eqn_Braiding}
\sigma_{\SpaceH \SpaceG} :=
\sum_{\varsigma(n,m)=1}^{\dim{\SpaceH}\dim{\SpaceG}}
\ket{f_m} \otimes \ket{e_n} \; \bra{e_n}\otimes \bra{f_m}.
\end{equation}

%% Dagger
\noindent Finally, one can define a dagger on morphisms by taking the conjugate transpose on the \textbf{matrix representation} given by (\ref{eqn_MatrixRepresentation}), obtaining the following morphism (by transfer theorem):
\begin{equation}\label{eqn_Dagger}
(\truncate{F})^\dagger := 
\sum_{n=1}^{\dim{\SpaceH}} 
\sum_{m=1}^{\dim{\SpaceG}} 
\ket{e_{n}} 
\truncate{F}_{mn}^\star
\bra{f_{m}}.
\end{equation}
It is straightforward to check that $(\truncate{F})^\dagger$ is a morphism $\SpaceG \rightarrow \SpaceH$ whenever $\truncate{F}$ is a morphism $\SpaceH \rightarrow \SpaceG$, that the dagger is functorial and that it satisfies all the compatibility requirements with the monoidal structure. The content of this section can thus be summarised by the following result.

\begin{theorem}
The category $\starHilbCategory$ is a $\dagger$-SMC, with tensor product and dagger defined by (\ref{eqn_TensorObjects}, \ref{eqn_TensorMorphisms}, \ref{eqn_Dagger}).
\end{theorem} 

%\newpage
\subsection{Standard bounded linear maps in \texorpdfstring{$\starHilbCategory$}{*Hilb}}
\label{section_StandardBoundedMaps}

%% Restriction to separable Hilbert spaces
In order to do categorical quantum mechanics in $\starHilbCategory$, we have to first establish its relationship with the more traditional arena of standard Hilbert spaces and bounded linear maps. By construction, we don't expect $\starHilbCategory$ to contain all of $\HilbCategory$, as the objects were explicitly chosen to be separable (rather than arbitrary) Hilbert spaces. We expect, however, that the full subcategory $\sHilbCategory$ of separable Hilbert spaces and bounded linear maps will be faithfully embedded in it. 

We will refer to morphisms $\ket{\psi} :\equiv \sum_{n=1}^{\dim{\SpaceH}} \psi_n \ket{e_n}: \starComplexs \rightarrow \SpaceH$ as \textbf{vectors} or \textbf{states} in $\SpaceH$, and the $\starComplexs$-valued inner product induced by the dagger can be written as $\braket{\phi}{\psi} = \sum_{n=1}^{\dim{\SpaceH}} \phi_n^\star \psi_n$. We will refer to vectors $\ket{\psi}$ having finite squared norm $\braket{\psi}{\psi}$ as \textbf{finite vectors}, and to vectors having infinitesimal squared norm as \textbf{infinitesimal vectors}. Difference by infinitesimal vectors gives rise to the following equivalence relation, corresponding to the notion of convergence of vectors in norm:
\begin{equation}
\ket{\phi} \simeq \ket{\psi} \iffdef \ket{\phi}-\ket{\psi} \text{ is infinitesimal}.
\end{equation}
We will say that a morphism $\truncate{F}: \SpaceH \rightarrow \SpaceG$ in $\starHilbCategory$ is \textbf{continuous} if for any $\ket{\psi_\kappa},\ket{\phi_\kappa} :\, \starComplexs \rightarrow \SpaceH$ satisfying $\ket{\psi_\kappa} \simeq \ket{\phi_\kappa}$  we have $\truncate{F} \ket{\psi_\kappa} \simeq \truncate{F} \ket{\phi_\kappa}$. Furthermore, the \textbf{operator norm} on some homset $\Hom{\starHilbCategory}{\SpaceH}{\SpaceG}$ can be defined as follows\footnote{Both the $\sup$ and the square root are simply extended from $\reals^+$ to $\starReals^+$ by transfer theorem, as usual. The definition of the operator norm is independent of the equivalence relation $\simeq$.}:
\begin{equation}
|| \truncate{F} ||_{op} := \sup_{\braket{\psi}{\psi} = 1} \sqrt{\bra{\psi}\truncate{F}^\dagger \truncate{F} \ket{\psi} }.
\end{equation}
We will say that a morphism $\truncate{F}: \SpaceH \rightarrow \SpaceG$ is \textbf{bounded} if its operator norm $||\truncate{F}||_{op}$ is finite. Just as it happens in the case of standard Hilbert spaces, throughout this work we will confuse bounded and continuous, thanks to the following result.

\begin{lemma}\label{thm_Continuity}
Let $\truncate{F}: \SpaceH \rightarrow \SpaceG$ be a morphism in $\starHilbCategory$. The following are equivalent:
\begin{enumerate} 
\item[(i)] the operator norm $||\truncate{F}||_{op}$ is finite;
\item[(ii)] $\truncate{F} \ket{\xi_\kappa}:\, \starComplexs \rightarrow \SpaceG$ is infinitesimal whenever $\ket{\xi_\kappa} :\, \starComplexs \rightarrow \SpaceH$ is infinitesimal;
\item[(iii)] if $\ket{\psi_\kappa},\ket{\phi_\kappa} :\, \starComplexs \rightarrow \SpaceH$ satisfy $\ket{\psi_\kappa} \simeq \ket{\phi_\kappa}$, then we have $\truncate{F} \ket{\psi_\kappa} \simeq \truncate{F} \ket{\phi_\kappa}$.
\end{enumerate}
\end{lemma}

\noindent The following equivalence relation embodies the notion of convergence in operator norm:
\begin{equation}\label{eqn_InfinitesimalOpEquiv}
\truncate{F} \sim \truncate{F'} \iffdef ||\truncate{F} - \truncate{F'}||_{op} \text{ is infinitesimal}.
\end{equation}
This equivalence relation is $\complexs$-linear, by triangle inequality, and it commutes with the dagger. It also commutes with composition and tensor product, as long as we restrict ourselves to continuous operators.
\begin{lemma}\label{thm_InfinitesimalOpEquiv}
Suppose that $\truncate{F}$, $\truncate{F'}$, $\truncate{G}$ and $\truncate{G'}$ are all continuous. Then the following statements hold:
\begin{align}
\truncate{G} \cdot \truncate{F} \sim \truncate{G'} \cdot \truncate{F'} \text{ whenever both } \truncate{F} \sim \truncate{F'} \text{ and } \truncate{G} \sim \truncate{G'}, \nonumber \\
\truncate{G} \otimes \truncate{F} \sim \truncate{G'} \otimes \truncate{F'} \text{ whenever both } \truncate{F} \sim \truncate{F'} \text{ and } \truncate{G} \sim \truncate{G'}. 
\end{align}
\end{lemma}

We say that a morphism $\truncate{G}$ is \textbf{near-standard} (in the operator norm) if it satisfies $\truncate{G} \sim \truncate{f}$ for some standard bounded linear map $f$. From now on, we will always use lowercase letters to denote standard bounded linear maps. Near-standard morphisms are in particular continuous, and form a sub-$\dagger$-SMC of $\starHilbCategory$, which we shall denote by $\starHilbCategoryNearStd$. By Lemma \ref{thm_InfinitesimalOpEquiv} this subcategory can be enriched to become a strict $\dagger$-symmetric monoidal 2-category, with 2-cells given by the identity 2-cell and the idempotent 2-cell $\sim$. This observation finally allows us to relate $\starHilbCategory$ and $\sHilbCategory$. 

\newpage
\noindent We define a strict \textbf{standard part} functor $\stdpart{\emptyArg}:\, \starHilbCategoryNearStd \rightarrow \sHilbCategory$ as follows:
\begin{enumerate}
\item[(i)] $\stdpart{V,\ket{e_{n}}_{n=1}^\kappa} := V$;
\item[(ii)] $\stdpart{\truncate{F}} := $ the unique $f'$ such that $f'$ is a standard bounded linear map and $\truncate{F} \sim \truncate{f'}$.
\end{enumerate}
\noindent We fix an infinite non-standard natural $\omega$, and define a weak functor $\vartruncate{\emptyArg}_\omega: \sHilbCategory \rightarrow\, \starHilbCategoryNearStd$ as follows (with functoriality only up to $\sim$):
\begin{enumerate}
\item[(i)] $\vartruncate{V}_\omega := (V,(\ket{e_{n}})_{n})$, where the orthonormal bases are chosen in such a way as to respect tensor product of $\starHilbCategoryNearStd$ (see the Appendix for more details about this choice);
\item[(ii)] $\vartruncate{f}_\omega := \truncate{f}$ on morphisms, and Lemma \ref{thm_InfinitesimalOpEquiv} guarantees $\vartruncate{ G \cdot F}_\omega \sim \vartruncate{G}_\omega \cdot \vartruncate{F}_\omega$.
\end{enumerate}

\begin{theorem}\label{thm_SeparableInStarHilb}
The following results relate $\starHilbCategoryNearStd$ and $\sHilbCategory$:
\begin{enumerate}
\item[(i)] $\stdpart{\emptyArg}$ is a strict full functor of $\dagger$-SMCs, which is surjective on objects; 
\item[(ii)] $\vartruncate{\emptyArg}_\omega$ is a weak faithful functor from a $\dagger$-SMC to a $\dagger$-symmetric monoidal 2-category, which is essentially surjective on objects; its restriction to the subcategory $\fdHilbCategory$ is strictly functorial;
\item[(iii)] $\stdpart{\vartruncate{f}_\omega} = f$, for all standard bounded morphisms $f$ of separable Hilbert spaces;
\item[(iv)] $\stdpart{\vartruncate{V}_\omega} = V$, for all objects $V$ of $\sHilbCategory$;
\item[(v)] For all objects $\SpaceH$ of $\starHilbCategoryNearStd$, there is a (unique) standard unitary $\truncate{u}_\SpaceH : \SpaceH \rightarrow \vartruncate{\stdpart{\SpaceH}}_\omega$ such that $\stdpart{\truncate{u}_\SpaceH} = \id{\stdpart{\SpaceH}}$.
\item[(vi)] $\truncate{u}_\SpaceG^\dagger \vartruncate{\stdpart{\truncate{F}}}_\omega \truncate{u}_\SpaceH \sim \truncate{F}$ for all morphisms $\truncate{F}: \SpaceH \rightarrow \SpaceG$ in $\starHilbCategoryNearStd$
\end{enumerate}
\end{theorem}

\noindent The essence of Theorem \ref{thm_SeparableInStarHilb} is that $\sHilbCategory$ is equivalent to the subcategory $\starHilbCategoryNearStd$ of $\starHilbCategory$ given by near-standard morphisms in the operator norm, as long as we take care to equate morphisms which are infinitesimally close. The equivalence allows one to prove results about $\sHilbCategory$ by working in $\starHilbCategory$ and taking advantage of the CQM machinery introduced in the next Section: in a typical scenario, one would (i) start from $\sHilbCategory$, (ii) lift to $\starHilbCategoryNearStd$ by using $\vartruncate{\emptyArg}_{\omega}$, (iii) work in $\starHilbCategory$, (iv) obtain a result in $\starHilbCategoryNearStd$, (v) descend to $\sHilbCategory$ again by using $\stdpart{\emptyArg}$. This is conceptually akin to using the two directions of the transfer theorem to prove results of standard analysis using non-standard methods.

However, one shouldn't necessarily discount $\starHilbCategory$ as just being a category of handy mathematical tricks: as we shall now proceed to see, a number of objects of concrete interest in the everyday practice of quantum mechanics (such as the position/momentum observables and eigenstates) are native to that richer environment, and confer it its own independent dignity.

\section{Infinite-dimensional categorical quantum mechanics}
\label{section_InfiniteDimCQM}

\subsection{Classical structures in \texorpdfstring{$\starHilbCategory$}{*Hilb}}
\label{section_ClassicalStructures}

Our main motivation comes from the work of \cite{abramsky2012hstaralgebras} on commutative $H^\star$-algebras, a particular class of non-unital special commutative $\dagger$-Frobenius algebras\footnote{In \cite{abramsky2012hstaralgebras}, non-unital special commutative $\dagger$-Frobenius algebras are simply referred to as \textit{Frobenius algebras}. In this work we will refer to them in full as special commutative $\dagger$-Frobenius algebras. We will specify \textit{non-unital} or \textit{unital} explicitly.} (non-unital $\dagger$-SCFAs, in short). It is an established result that approximate units for the algebras exist in separable Hilbert spaces: we will show that, in our non-standard framework, they can be made truly unital.

\begin{theorem}[From \cite{abramsky2012hstaralgebras}]
A non-unital $\dagger$-SCFA $(\hbox{\begin{tikzpicture} [scale=0.8,transform shape] %% DO NOT CHANGE

\def\deltax{0.3} %% CAN BE CHANGED
\def\deltay{0.5} %% DO NOT CHANGE

%\path[use as bounding box] (-\deltax,-\deltay) rectangle (\deltax,\deltay);

\node (mult_label_outl) at (-\deltax,+\deltay) {};
\node (mult_label_outr) at (+\deltax,+\deltay) {};
\node [dot, fill=\Zbwcolour] (mult) at (0,0) {};
\node (mult_label_in) at (0,-\deltay) {};
\draw[-] [in=270,out=135] (mult) to (mult_label_outl);
\draw[-] [in=270,out=45] (mult) to (mult_label_outr);
\draw[-] (mult_label_in) to (mult);

%\draw (current bounding box.south west) rectangle (current bounding box.north east);
\end{tikzpicture}}\!,\hbox{\begin{tikzpicture} [scale=0.8,transform shape] %% DO NOT CHANGE

\def\deltax{0.3} %% CAN BE CHANGED
\def\deltay{0.5} %% DO NOT CHANGE

%\path[use as bounding box] (-\deltax,-\deltay) rectangle (\deltax,\deltay);

\node (mult_label_inl) at (-\deltax,-\deltay) {};
\node (mult_label_inr) at (+\deltax,-\deltay) {};
\node [dot, fill=\Zbwcolour] (mult) at (0,0) {};
\node (mult_label_out) at (0,+\deltay) {};

\draw[-] [out=90,in=225](mult_label_inl) to (mult);
\draw[-] [out=90,in=315](mult_label_inr) to (mult);
\draw[-] (mult) to (mult_label_out);

%\draw (current bounding box.south west) rectangle (current bounding box.north east);
\end{tikzpicture}}\!\!)$ on an object $V$ of $\sHilbCategory$ is an $H^\star$-algebra if and only if it corresponds to an orthonormal basis $\ket{e_{n}}_{n\in \naturals^+}$ of $V$ such that $\hbox{\begin{tikzpicture} [scale=0.8,transform shape] %% DO NOT CHANGE

\def\deltax{0.3} %% CAN BE CHANGED
\def\deltay{0.5} %% DO NOT CHANGE

%\path[use as bounding box] (-\deltax,-\deltay) rectangle (\deltax,\deltay);

\node (mult_label_outl) at (-\deltax,+\deltay) {};
\node (mult_label_outr) at (+\deltax,+\deltay) {};
\node [dot, fill=\Zbwcolour] (mult) at (0,0) {};
\node (mult_label_in) at (0,-\deltay) {};
\draw[-] [in=270,out=135] (mult) to (mult_label_outl);
\draw[-] [in=270,out=45] (mult) to (mult_label_outr);
\draw[-] (mult_label_in) to (mult);

%\draw (current bounding box.south west) rectangle (current bounding box.north east);
\end{tikzpicture}}\! \cdot \ket{e_n} = \ket{e_n}\ket{e_n}$.
\end{theorem}

\newpage
\begin{theorem}[From \cite{abramsky2012hstaralgebras,ambrose1945structure}]
A $\dagger$-SCFA $(\hbox{\begin{tikzpicture} [scale=0.8,transform shape] %% DO NOT CHANGE

\def\deltax{0.3} %% CAN BE CHANGED
\def\deltay{0.5} %% DO NOT CHANGE

%\path[use as bounding box] (-\deltax,-\deltay) rectangle (\deltax,\deltay);

\node (mult_label_outl) at (-\deltax,+\deltay) {};
\node (mult_label_outr) at (+\deltax,+\deltay) {};
\node [dot, fill=\Zbwcolour] (mult) at (0,0) {};
\node (mult_label_in) at (0,-\deltay) {};
\draw[-] [in=270,out=135] (mult) to (mult_label_outl);
\draw[-] [in=270,out=45] (mult) to (mult_label_outr);
\draw[-] (mult_label_in) to (mult);

%\draw (current bounding box.south west) rectangle (current bounding box.north east);
\end{tikzpicture}}\!,\hbox{\begin{tikzpicture} [scale=0.8,transform shape] %% DO NOT CHANGE

\def\deltax{0.3} %% CAN BE CHANGED
\def\deltay{0.5} %% DO NOT CHANGE

%\path[use as bounding box] (-\deltax,-\deltay) rectangle (\deltax,\deltay);

\node (mult_label_inl) at (-\deltax,-\deltay) {};
\node (mult_label_inr) at (+\deltax,-\deltay) {};
\node [dot, fill=\Zbwcolour] (mult) at (0,0) {};
\node (mult_label_out) at (0,+\deltay) {};

\draw[-] [out=90,in=225](mult_label_inl) to (mult);
\draw[-] [out=90,in=315](mult_label_inr) to (mult);
\draw[-] (mult) to (mult_label_out);

%\draw (current bounding box.south west) rectangle (current bounding box.north east);
\end{tikzpicture}}\!\!)$ on an object $V$ of $\sHilbCategory$ is an $H^\star$-algebra if and only if there is a sequence $\ket{E_n}_{n \in \naturals^+}$ such that for all $\ket{a} : \complexs \rightarrow V$ we have:
\begin{enumerate}
\item[(i)] $\hbox{\begin{tikzpicture} [scale=0.8,transform shape] %% DO NOT CHANGE

\def\deltax{0.3} %% CAN BE CHANGED
\def\deltay{0.5} %% DO NOT CHANGE

%\path[use as bounding box] (-\deltax,-\deltay) rectangle (\deltax,\deltay);

\node (mult_label_inl) at (-\deltax,-\deltay) {};
\node (mult_label_inr) at (+\deltax,-\deltay) {};
\node [dot, fill=\Zbwcolour] (mult) at (0,0) {};
\node (mult_label_out) at (0,+\deltay) {};

\draw[-] [out=90,in=225](mult_label_inl) to (mult);
\draw[-] [out=90,in=315](mult_label_inr) to (mult);
\draw[-] (mult) to (mult_label_out);

%\draw (current bounding box.south west) rectangle (current bounding box.north east);
\end{tikzpicture}}\! \cdot (\ket{E_n} \otimes \ket{a})$ converges to $\ket{a}$;
\item[(ii)] $(\id{V} \otimes \bra{a}) \cdot \hbox{\begin{tikzpicture} [scale=0.8,transform shape] %% DO NOT CHANGE

\def\deltax{0.3} %% CAN BE CHANGED
\def\deltay{0.5} %% DO NOT CHANGE

%\path[use as bounding box] (-\deltax,-\deltay) rectangle (\deltax,\deltay);

\node (mult_label_outl) at (-\deltax,+\deltay) {};
\node (mult_label_outr) at (+\deltax,+\deltay) {};
\node [dot, fill=\Zbwcolour] (mult) at (0,0) {};
\node (mult_label_in) at (0,-\deltay) {};
\draw[-] [in=270,out=135] (mult) to (mult_label_outl);
\draw[-] [in=270,out=45] (mult) to (mult_label_outr);
\draw[-] (mult_label_in) to (mult);

%\draw (current bounding box.south west) rectangle (current bounding box.north east);
\end{tikzpicture}}\! \cdot \ket{E_n}$ converges.
\end{enumerate}
If this is so, then we can take $\ket{E_{n}} := \sum_{n'\leq n} \ket{e_{n'}}$.
\end{theorem}

\noindent The sequence $\ket{E_{n}}_{n \in \naturals^+}$ itself doesn't converge in $\sHilbCategory$, because the state $\sum_{n \in \naturals^+} \ket{e_{n}}$ would have infinite norm. In our non-standard context, however, the state $\sum_{n=1}^{\kappa} \ket{e_n}$ is a well-defined, internal state for $\SpaceH = (V,\ket{e_n}_{n=1}^\kappa)$. This opens the way to the definition of unital $\dagger$-SCFAs on all objects of $\starHilbCategory$.

\begin{theorem}\label{thm_ClassicalStructures}
Let $\SpaceH = (V,\ket{e_n}_{n=1}^\kappa)$ be an object in $\starHilbCategory$, and $\ket{f_{n}}_{n \in \naturals^+}$ be a standard orthonormal basis for $V$. Then the following comultiplication and counit define a \textbf{weakly unital}, \textbf{weakly special} commutative $\dagger$-Frobenius algebra on $\SpaceH$ (i.e. one where the Unit and Speciality laws hold only up to $\sim$):
\begin{equation}
	\begin{multlined}
	\begin{tikzpicture}[node distance = 0.5cm]
		\node[dot,fill=\Zbwcolour] (comultdot) {};
		\node (comultin) [below of = comultdot] {};
		\node (comultoutl) [above of = comultdot, xshift = -0.5cm] {}; 
		\node (comultoutr) [above of = comultdot, xshift = +0.5cm] {};
		\begin{pgfonlayer}{background}
		\draw[-] (comultin.270) to (comultdot);
		\draw[-,out=135,in=270] (comultdot.135) to (comultoutl.90);
		\draw[-,out=45,in=270] (comultdot.45) to (comultoutr.90);
		\end{pgfonlayer}
		\node (comultdef) [right of = comultdot, xshift = 1.5cm] {$:= \sum\limits_{n=1}^\kappa \ket{f_n}\otimes\ket{f_n}\;\bra{f_n}$};
		\node[dot,fill=\Zbwcolour] (counitdot) [right of = comultdef, xshift = 3cm]{};
		\node (counitin) [below of = counitdot] {};
		\begin{pgfonlayer}{background}
		\draw[-] (counitin.270) to (counitdot);
		\end{pgfonlayer}
		\node (counitdef) [right of = counitdot, xshift = 0.75cm] {$:= \sum\limits_{n=1}^\kappa \bra{f_n}$};
	\end{tikzpicture}
	\end{multlined}
\end{equation}
We refer to it as the \textbf{classical structure}\footnote{The terminology \textit{classical structure}, in the context of $\starHilbCategory$, will refer to weakly unital, weakly special, commutative $\dagger$-Frobenius algebras. This is in accordance with the weak functoriality of $\vartruncate{\emptyArg}_\omega$ seen in the previous section.} 
for $\ket{f_n}_n$. When $\ket{f_n}_{n}$ is  the \textbf{chosen orthonormal basis} $\ket{e_n}_n$ for $\SpaceH$, the algebra is strictly unital and strictly special, i.e. a unital $\dagger$-SCFA.
\end{theorem}

\noindent The existence of unital $\dagger$-SCFAs makes $\starHilbCategory$ a dagger compact category, with self-dual objects.
\begin{theorem}\label{thm_CompactClosed}
The category $\starHilbCategory$ is compact closed, with \textbf{cap} and \textbf{cup} on an object $\SpaceH = (V,\ket{e_n}_{n=1}^\kappa)$ derived from the classical structure for the chosen orthonormal basis $\ket{e_n}_n$:
\begin{equation}
	\begin{multlined}
	\begin{tikzpicture}[node distance = 0.5cm]
		% Cap
		\node (capcenter) {};
		\node (capinl) [below of = capcenter, yshift = 0.15cm, xshift = -0.5cm] {};
		\node (capinr) [below of = capcenter, yshift = 0.15cm, xshift = +0.5cm] {};	
		\node (capoutl) [above of = capcenter, yshift = -0.5cm, xshift = -0.5cm] {};
		\node (capoutr) [above of = capcenter, yshift = -0.5cm, xshift = +0.5cm] {};	
		\begin{pgfonlayer}{background}
		\draw[-] (capinl.270) to (capoutl.90);
		\draw[-] (capinr.270) to (capoutr.90);
		\draw[-,out=90,in=90] (capoutl) to (capoutr);
		\end{pgfonlayer}
		\node (capdef) [right of = capcenter, xshift = 0.5cm] {$:=$};
		\node[dot,fill=\Zbwcolour] (multcenter) [right of = capdef, xshift = 0.5cm]{};
		\node (multinl) [below of = multcenter, yshift = 0.15cm, xshift = -0.5cm] {};
		\node (multinr) [below of = multcenter, yshift = 0.15cm, xshift = +0.5cm] {};
		\node[dot,fill=\Zbwcolour] (counit) [above of = multcenter]	{};		
		\begin{pgfonlayer}{background}
		\draw[-,in=225,out=90] (multinl.270) to (multcenter);
		\draw[-,in=315,out=90] (multinr.270) to (multcenter);
		\draw[-,out=90,in=270] (multcenter) to (counit);
		\end{pgfonlayer}
		\node (capsum) [right of = multcenter, xshift = 1.25cm] {$= \sum\limits_{n=1}^{\kappa} \bra{e_n}\otimes\bra{e_n}$};
		% Cup
		\node (capcenter) [right of = capsum, xshift = 3cm]{};
		\node (capinl) [below of = capcenter, yshift = 0.5cm, xshift = -0.5cm] {};
		\node (capinr) [below of = capcenter, yshift = 0.5cm, xshift = +0.5cm] {};	
		\node (capoutl) [above of = capcenter, yshift = -0.15cm, xshift = -0.5cm] {};
		\node (capoutr) [above of = capcenter, yshift = -0.15cm, xshift = +0.5cm] {};	
		\begin{pgfonlayer}{background}
		\draw[-] (capinl.270) to (capoutl.90);
		\draw[-] (capinr.270) to (capoutr.90);
		\draw[-,out=270,in=270] (capinl) to (capinr);
		\end{pgfonlayer}
		\node (capdef) [right of = capcenter, xshift = 0.5cm] {$:=$};
		\node[dot,fill=\Zbwcolour] (multcenter) [right of = capdef, xshift = 0.5cm]{};
		\node (multinl) [above of = multcenter, yshift = -0.15cm, xshift = -0.5cm] {};
		\node (multinr) [above of = multcenter, yshift = -0.15cm, xshift = +0.5cm] {};
		\node[dot,fill=\Zbwcolour] (counit) [below of = multcenter]	{};		
		\begin{pgfonlayer}{background}
		\draw[-,in=135,out=270] (multinl.90) to (multcenter);
		\draw[-,in=45,out=270] (multinr.90) to (multcenter);
		\draw[-,out=270,in=90] (multcenter) to (counit);
		\end{pgfonlayer}
		\node (capsum) [right of = multcenter, xshift = 1.25cm] {$= \sum\limits_{n=1}^{\kappa} \ket{e_n}\otimes\ket{e_n}$};
	\end{tikzpicture}
	\end{multlined}
\end{equation}
More in general, any classical structure in $\starHilbCategory$ can be used to define a \textbf{weak cap} and a \textbf{weak cup}, satisfying weak yanking equations (i.e. with equality only up to $\sim$).
\end{theorem}

\noindent The compact closed structure gives rise to a \textbf{trace}  in the usual way:
\begin{equation}
	\begin{multlined}
	\begin{tikzpicture}[node distance = 0.5cm]
		\node (tr) {$\Trace{\truncate{F}} := $};
		\node[box] (F) [right of = tr, xshift = 0.6cm]{$\truncate{F}$};
		\node (in) [below of = F]{};
		\node (inr) [right of = in]{};
		\node (out) [above of = F]{};
		\node (outr) [right of = out]{};	
		\begin{pgfonlayer}{background}
		\draw[-] (in) to (F);
		\draw[-] (F) to (out);
		\draw[-,out=90,in=90] (out.270) to (outr.270);
		\draw[-,out=270,in=270] (in.90) to (inr.90);
		\draw[-] (inr) to (outr);
		\end{pgfonlayer}
		\node (trdef) [right of = F, xshift = 1cm] {$=\sum\limits_{n=1}^{\dim{\SpaceH}} \truncate{F}_{nn}$};
		\node (tr) [right of = trdef, xshift = 2.5cm] {$\pTrace{\truncate{G}}{\SpaceH}{\SpaceG}{\SpaceK} := $};
		\node[box, minimum width = 0.75cm] (F) [right of = tr, xshift = 1cm]{$\truncate{G}$};
		\node (in) [below of = F, xshift = 0.2cm]{};
		\node (inl) [below of = F, xshift = -0.2cm,yshift = -0.25cm]{$\SpaceH$};
		\node (inr) [right of = in]{};
		\node (out) [above of = F, xshift = 0.2cm]{};
		\node (outl) [above of = F, xshift = -0.2cm,yshift = 0.25cm]{$\SpaceG$};	
		\node (outr) [right of = out]{};
		\node (labelK) [right of = inr, xshift = -0.3cm,yshift = 0cm] {$\SpaceK$};	
		\begin{pgfonlayer}{background}
		\draw[-] (in) to (out);
		\draw[-,out=90,in=90] (out.270) to (outr.270);
		\draw[-,out=270,in=270] (in.90) to (inr.90);
		\draw[-] (inr) to (outr);
		\draw[-] (inl) to (outl);
		\end{pgfonlayer}
		\node (trdef) [right of = F, xshift = 3.4cm] {$=\sum\limits_{m=1}^{\dim{\SpaceG}}\sum\limits_{n=1}^{\dim{\SpaceH}}\sum\limits_{k=1}^{\dim{\SpaceK}} \ket{f_m}\truncate{G}_{\varsigma(m,k)\varsigma(n,k)}\bra{e_n}$};
	\end{tikzpicture}
	\end{multlined}
\end{equation}

\noindent In particular, we see that the notation $\dim{\SpaceH} := \kappa$ for $\SpaceH = (V,\ket{e_n}_{n=1}^\kappa)$ was well chosen: $\Trace{\id{\SpaceH}} = \sum_{n=1}^\kappa 1 = \kappa = \dim{\SpaceH}$. The trace can also be used to endow the homset $\Hom{\starHilbCategory}{\SpaceH}{\SpaceG}$, which we've already seen to be a $\starComplexs$-vector space, with the following $\starComplexs$-valued \textbf{Hilbert-Schmidt inner product}:
\begin{equation}
\innerprod{\truncate{G}}{\truncate{F}} := \Trace{\truncate{G}^\dagger \truncate{F}} = \sum_{m=1}^{\dim{\SpaceG}} \sum_{n=1}^{\dim{\SpaceH}} \truncate{G}_{mn}^\star \truncate{F}_{mn}.
\end{equation}
This is exactly the inner product that one would get by enriching the category $\starHilbCategory$ in itself via compact closure.

\newpage
\subsection{Wavefunctions with periodic boundaries}
\label{section_WavefunctionsPeriodic}

As a sample application of the structures presented above, we cover the theory of wavefunctions on a 1-dimensional space with periodic boundary conditions: these live in $\Ltwo{\reals/(L \integers)} \isom \Ltwo{S^1}$, where $L$ is the length of the underlying space. The \textbf{momentum eigenstates}, or \textbf{plane-waves}, form a countable orthogonal basis for $\Ltwo{\reals/(L \integers)}$, indexed by $n \in \integers$ (in this section, $n,m,k,h$ will range over integers):
\begin{equation}
\chi_n :=  x \mapsto  e^{-i (2\pi/L)nx}.
\end{equation}
The plane-wave $\ket{\chi_n}$ is the eigenstate of momentum $n \hbar$. Let $\theta(n) := |2n| + \frac{1-\operatorname{sign}(n)}{2}$ (with $\operatorname{sign}(0):=-1$) be a bijection $\integers \rightarrow \naturals^+$. We can obtain a countable orthonormal basis $\ket{e_{l}}_{l \in \naturals^+}$ for $\Ltwo{\reals/(L\integers)}$ as follows: 
\begin{equation}
\ket{e_{\theta(n)}} := \frac{1}{\sqrt{L}} \ket{\chi_{n}} \text{ for all } n \in \integers.
\end{equation}
Now we shift our attention to the object $(\Ltwo{\reals/(L\integers)}, \ket{e_l}_{l=1}^{\kappa})$ of $\starHilbCategory$, with $\kappa = 2 \omega + 1$ some odd infinite natural\footnote{The notions of oddness and evenness extend from $\naturals$ to $\starNaturals$ by transfer theorem, and by saying that some infinite non-standard natural $\kappa \in \starNaturals$ is odd we mean exactly that $\kappa = 2 \omega + 1$ for some (necessarily infinite) non-standard natural $\omega \in \starNaturals$. Note that the infinite natural $\omega$ here has nothing to do with the ordinal $\omega$ from set theory.}. As a shorthand for $\sum_{l=1}^{\kappa} \ket{\chi_{\theta^{-1}(l)}}$, and other cases where the index is bijected to the integers, we will simply re-index over the non-standard integers $\{-\omega,...,+\omega\}$ (such as in $\sum_{n=-\omega}^{+\omega} \ket{\chi_n}$). In particular, we will write our chosen object as $(\Ltwo{\reals/(L\integers)},\frac{1}{\sqrt{L}}\ket{\chi_n}_{n=-\omega}^{+\omega})$, or simply $\Ltwo{\reals/(L\integers)}$ \vspace{-1mm} when no confusion can arise. Now that we established the role of momentum eigenstates in our framework, it's time to turn our attention to position eigenstates. On a continuous space, position eigenstates are given by Dirac delta functions, and as a consequence are not associated with well-defined standard vectors. Here, we will define them in terms of the basis of momentum eigenstates, as follows, and then show that they coincide with their more traditional formulation in terms of Dirac deltas. Let $x_0 \in \reals/(L\integers)$ be a standard point of the underlying space, then we define the \textbf{position eigenstate} at $x_0$ to be the following non-standard state:
\begin{equation}
\ket{\delta_{x_0}} := \frac{1}{\sqrt{L}}\sum_{n=-\omega}^{+\omega}  \chi_{n}(x_0)^\star \frac{1}{\sqrt{L}}\ket{\chi_n}.
\end{equation}  

\begin{theorem}\label{thm_PositionEigenstates}
The position eigenstates are orthogonal (up to infinitesimals). Furthermore, they behave as \textbf{Dirac deltas}, i.e. they satisfy $\braket{\delta_{x_0}}{f} \simeq f(0)$ for all standard smooth $f \in \Ltwo{\reals/(L\integers)}$. The position eigenstates are also unbiased with respect to the momentum eigenstates, in the sense that $|\braket{\delta_{x_0}}{\chi_n}|=1$ independently of $n$ or $x_0$.
\end{theorem} 

Mutual unbias (aka complementarity) is already sufficient to provide a first approximation of the usual position/momentum uncertainty principle: it can be used to show that measuring any state in the position observable and then in the momentum observable (or vice versa) always yields the totally mixed state. We sketch the argument informally in the appendix, and we leave further treatment of the connection between complementarity and the uncertainty principle (in its various versions) to future work.

In the traditional setting of $\Ltwo{\reals}$, momenta are the infinitesimal generators of space translations:
\begin{equation}\label{eqn_MomentaGenerateTranslationsTraditional}
	\ket{x + y} = \exp[i\frac{x \textbf{p}}{\hbar}] \ket{y}.
\end{equation}
The following theorem shows the relationship between momentum eigenstates and position-space translation in our framework, and relates it to the special case of Equation \ref{eqn_MomentaGenerateTranslationsTraditional}. This more general relationship can be extended to cases (such as the finite-dimensional ones of \cite{gogioso2015categorical}) where momenta cannot be valued in a subset of the reals.

\newpage
\begin{theorem}\label{thm_MomentaGenerateTranslations}
Let $(\hbox{\begin{tikzpicture} [scale=0.8,transform shape] %% DO NOT CHANGE

\def\deltax{0.3} %% CAN BE CHANGED
\def\deltay{0.5} %% DO NOT CHANGE

%\path[use as bounding box] (-\deltax,-\deltay) rectangle (\deltax,\deltay);

\node (mult_label_outl) at (-\deltax,+\deltay) {};
\node (mult_label_outr) at (+\deltax,+\deltay) {};
\node [dot, fill=\Zbwcolour] (mult) at (0,0) {};
\node (mult_label_in) at (0,-\deltay) {};
\draw[-] [in=270,out=135] (mult) to (mult_label_outl);
\draw[-] [in=270,out=45] (mult) to (mult_label_outr);
\draw[-] (mult_label_in) to (mult);

%\draw (current bounding box.south west) rectangle (current bounding box.north east);
\end{tikzpicture}}\!,\hbox{\begin{tikzpicture} [scale=0.8,transform shape] %% DO NOT CHANGE

\def\deltax{0.3} %% CAN BE CHANGED
\def\deltay{0.5} %% DO NOT CHANGE

\path[use as bounding box] (-\deltax,-\deltay) rectangle (\deltax,\deltay);

\node [dot, fill=\Zbwcolour] (mult) at (0,0.25*\deltay) {};
\node (mult_label_in) at (0,-\deltay) {};
\draw[-] (mult_label_in) to (mult);

%\draw (current bounding box.south west) rectangle (current bounding box.north east);
\end{tikzpicture}}\!,\hbox{\begin{tikzpicture} [scale=0.8,transform shape] %% DO NOT CHANGE

\def\deltax{0.3} %% CAN BE CHANGED
\def\deltay{0.5} %% DO NOT CHANGE

%\path[use as bounding box] (-\deltax,-\deltay) rectangle (\deltax,\deltay);

\node (mult_label_inl) at (-\deltax,-\deltay) {};
\node (mult_label_inr) at (+\deltax,-\deltay) {};
\node [dot, fill=\Zbwcolour] (mult) at (0,0) {};
\node (mult_label_out) at (0,+\deltay) {};

\draw[-] [out=90,in=225](mult_label_inl) to (mult);
\draw[-] [out=90,in=315](mult_label_inr) to (mult);
\draw[-] (mult) to (mult_label_out);

%\draw (current bounding box.south west) rectangle (current bounding box.north east);
\end{tikzpicture}}\!,\hbox{\begin{tikzpicture} [scale=0.8,transform shape] %% DO NOT CHANGE

\def\deltax{0.3} %% CAN BE CHANGED
\def\deltay{0.5} %% DO NOT CHANGE

\path[use as bounding box] (-\deltax,-\deltay) rectangle (\deltax,\deltay);

\node [dot, fill=\Zbwcolour] (mult) at (0,-0.25*\deltay) {};
\node (mult_label_out) at (0,+\deltay) {};
\draw[-] (mult) to (mult_label_out);

%\draw (current bounding box.south west) rectangle (current bounding box.north east);
\end{tikzpicture}}\!)$ be the classical structure for the chosen orthonormal basis of normalised momentum eigenstates. Then the monoid $(\hbox{\begin{tikzpicture} [scale=0.8,transform shape] %% DO NOT CHANGE

\def\deltax{0.3} %% CAN BE CHANGED
\def\deltay{0.5} %% DO NOT CHANGE

%\path[use as bounding box] (-\deltax,-\deltay) rectangle (\deltax,\deltay);

\node (mult_label_inl) at (-\deltax,-\deltay) {};
\node (mult_label_inr) at (+\deltax,-\deltay) {};
\node [dot, fill=\Zbwcolour] (mult) at (0,0) {};
\node (mult_label_out) at (0,+\deltay) {};

\draw[-] [out=90,in=225](mult_label_inl) to (mult);
\draw[-] [out=90,in=315](mult_label_inr) to (mult);
\draw[-] (mult) to (mult_label_out);

%\draw (current bounding box.south west) rectangle (current bounding box.north east);
\end{tikzpicture}}\!,\hbox{\begin{tikzpicture} [scale=0.8,transform shape] %% DO NOT CHANGE

\def\deltax{0.3} %% CAN BE CHANGED
\def\deltay{0.5} %% DO NOT CHANGE

\path[use as bounding box] (-\deltax,-\deltay) rectangle (\deltax,\deltay);

\node [dot, fill=\Zbwcolour] (mult) at (0,-0.25*\deltay) {};
\node (mult_label_out) at (0,+\deltay) {};
\draw[-] (mult) to (mult_label_out);

%\draw (current bounding box.south west) rectangle (current bounding box.north east);
\end{tikzpicture}}\!)$ endows the set $\suchthat{[\sqrt{L}\ket{\delta_x}]_\simeq}{x \in \reals/(L\integers)}$ of position eigenstates with the abelian group structure of position-space translation $(\reals/(L\integers),\oplus,0)$:
\begin{equation}\label{eqn_MomentaGenerateTranslations}
	\begin{multlined}
	\begin{tikzpicture}[node distance = 1cm]
		\node (center) {};
		\node[point, inner sep = -0.35mm] (state) [below of = center] {$\delta_{x \oplus y}$};
		\node (factor) [left of = state] {$\sqrt{L}$};
		\node (out) [above of = center] {};
		\node (capdef) [right of = center, xshift = 0.25cm] {$ = $};
		\node[dot,fill=\Zbwcolour] (multcenter) [right of = capdef, xshift = 1cm]{};
		\node[point,inner sep = 0.5mm] (multinl) [below of = multcenter, yshift = 0cm, xshift = -0.75cm] {$\delta_x$};
		\node (factor) [left of = multinl,xshift = 2mm] {$\sqrt{L}$};
		\node[point,inner sep = 0.35mm] (multinr) [below of = multcenter, yshift = 0cm, xshift = +0.75cm] {$\delta_y$};
		\node (factor) [right of = multinr, xshift = -4mm] {$\sqrt{L}$};
		\node (counit) [above of = multcenter]	{};		
		\begin{pgfonlayer}{background}
		\draw[-] (state) to (out);
		\draw[-,in=225,out=90] (multinl) to (multcenter);
		\draw[-,in=315,out=90] (multinr) to (multcenter);
		\draw[-,out=90,in=270] (multcenter) to (counit);
		\end{pgfonlayer}
	\end{tikzpicture}
	\end{multlined}
\end{equation}
where the compact abelian group structure $(\Ltwo{\reals/(L\integers)},\oplus,0)$ is given by real addition modulo $L \in \reals$. This can be equivalently written in the following form, drawing an explicit parallelism with Equation (\ref{eqn_MomentaGenerateTranslationsTraditional}):
\begin{equation}\label{eqn_MomentaGenerateTranslationsExplicit}
\ket{\delta_{x \oplus y}} = \Big[\frac{1}{\sqrt{L}^2}\sum_{n=-\omega}^{+\omega} \chi_{n}(x)^\star \ket{\chi_n}\bra{\chi_n} \Big] \ket{\delta_y},
\end{equation}
where $\chi_{n}(x)^\star \simeq \braket{\chi_n}{\delta_x} $ is nothing but $ \exp[i \frac{x\,p}{\hbar}]$, for a given (quantised) momentum eigenvalue $p = n \hbar$ and corresponding momentum eigenstate $\ket{\chi_n}$. As a consequence, we will refer to this fact by saying that \textbf{momenta generate position-space translation}.
\end{theorem}

In quantum foundations, observables are often identified with complete families of orthogonal projectors, rather then with self-adjoint operators\footnote{Complete families may be indexed by any set, while self-adjoint operators require indices to be real numbers. This is not suitable, for example, for all those applications (such as those of \cite{gogioso2015categorical}) where momenta/positions are not naturally valued in a subset of the reals.}. In our case, the \textbf{momentum observable} is simply the family $(P_{n \hbar})_{n\hbar}$ of orthogonal projectors on $\Ltwo{\reals/(L\integers)}$ indexed by the set of quantised momenta $\suchthat{n \hbar}{n \in \integers}$, where the projectors are 1-dimensional and given by $P_{n \hbar} := \ket{\chi_n}\bra{\chi_n}$. In categorical quantum mechanics, on the other hand, observables are identified with special $\dagger$-Frobenius algebras: the following theorem connects the two definitions.

\begin{theorem}\label{thm_MomentumObservable}
The momentum observable can be obtained from the comultiplication $\hbox{\begin{tikzpicture} [scale=0.8,transform shape] %% DO NOT CHANGE

\def\deltax{0.3} %% CAN BE CHANGED
\def\deltay{0.5} %% DO NOT CHANGE

%\path[use as bounding box] (-\deltax,-\deltay) rectangle (\deltax,\deltay);

\node (mult_label_outl) at (-\deltax,+\deltay) {};
\node (mult_label_outr) at (+\deltax,+\deltay) {};
\node [dot, fill=\Zbwcolour] (mult) at (0,0) {};
\node (mult_label_in) at (0,-\deltay) {};
\draw[-] [in=270,out=135] (mult) to (mult_label_outl);
\draw[-] [in=270,out=45] (mult) to (mult_label_outr);
\draw[-] (mult_label_in) to (mult);

%\draw (current bounding box.south west) rectangle (current bounding box.north east);
\end{tikzpicture}}\!$ of the classical structure for momentum eigenstates as follows:
\begin{equation}\label{eqn_MomentumObservable}
	\begin{multlined}
	\begin{tikzpicture}[node distance = 1cm]
		\node (center) {$P_{n\hbar}$};
		\node (capdef) [right of = center, xshift = -0.25cm] {$ = $};
		\node[dot,fill=\Zbwcolour] (multcenter) [right of = capdef, xshift = 0cm]{};
		\node (multinl) [above of = multcenter, yshift = 0cm, xshift = -0cm] {};
		\node[kpointdag,inner sep = 0.1mm] (multinr) [above of = multcenter, yshift = -0.25cm, xshift = +1cm] {$\chi_n$};
		\node (label) [right of = multinr,xshift = -3mm]{$\frac{1}{\sqrt{L}}$};
		\node (counit) [below of = multcenter]	{};		
		\begin{pgfonlayer}{background}
		\draw[-,in=90,out=270] (multinl.90) to (multcenter);
		\draw[-,in=45,out=270] (multinr) to (multcenter);
		\draw[-,out=270,in=90] (multcenter) to (counit);
		\end{pgfonlayer}
	\end{tikzpicture}
	\end{multlined}
\end{equation}
\end{theorem}

The definition of a position observable is not as straightforward, because the position eigenstates don't form a \textit{countable} orthonormal basis. This inconvenience, however, allows us to make an interesting detour. The momentum observable is given by the comultiplication $\hbox{\begin{tikzpicture} [scale=0.8,transform shape] %% DO NOT CHANGE

\def\deltax{0.3} %% CAN BE CHANGED
\def\deltay{0.5} %% DO NOT CHANGE

%\path[use as bounding box] (-\deltax,-\deltay) rectangle (\deltax,\deltay);

\node (mult_label_outl) at (-\deltax,+\deltay) {};
\node (mult_label_outr) at (+\deltax,+\deltay) {};
\node [dot, fill=\Zbwcolour] (mult) at (0,0) {};
\node (mult_label_in) at (0,-\deltay) {};
\draw[-] [in=270,out=135] (mult) to (mult_label_outl);
\draw[-] [in=270,out=45] (mult) to (mult_label_outr);
\draw[-] (mult_label_in) to (mult);

%\draw (current bounding box.south west) rectangle (current bounding box.north east);
\end{tikzpicture}}\!$, and its adjoint acts as the compact abelian group $\reals / (L \integers)$ on the set of (equivalence classes under $\simeq$ of) position eigenstates. This latter statement corresponds to momenta generating position-space translations (in terms of Fourier theory, the Pontryagin dual of the compact abelian group $\reals / (L \integers)$ is the discrete abelian group $\integers$). Dually, we know that positions should generate momentum-space translations (because $\reals / (L \integers)$ is Pontryagin dual to $\integers$), and hence we expect the adjoint of the position observable to act as the discrete abelian group $\integers$ on the space of momenta. Thus, in order to obtain the position observable, we first define the desired group action on the chosen orthonormal basis of momentum eigenstates.

\newpage
Consider the binary function $a \oplus b := \modclass{a+b}{2N+1}$, where representatives for the $2N+1$ remainder classes are chosen in the set $\{-N,...,+N\}$: for every $N$, this function is defined in the standard theory of $\integers$, and endows $\{-N,...,+N\}$ with the group structure of $\integersMod{2N+1}$. By transfer theorem, a similar group operation exists on the internal set $\{-\omega, ..., +\omega\}$ of $\starIntegers$, endowing it with the group structure of $\integersMod{2\omega + 1}$. Remarkably, for any two finite integers $n,m \in \integers$ we have $n \oplus m = n + m$ (because $n,m < \omega$ implies $n+m < \omega$, so no modular reduction occurs). Now consider the following morphisms of $\starHilbCategory$:
\begin{equation}
	\begin{multlined}
	\begin{tikzpicture}[node distance = 0.5cm]
		\node[dot,fill=\Xbwcolour] (comultdot) {};
		\node (comultin) [above of = comultdot] {};
		\node (comultoutl) [below of = comultdot, xshift = -0.5cm] {}; 
		\node (comultoutr) [below of = comultdot, xshift = +0.5cm] {};
		\begin{pgfonlayer}{background}
		\draw[-] (comultin.90) to (comultdot);
		\draw[-,out=225,in=90] (comultdot.225) to (comultoutl.270);
		\draw[-,out=315,in=90] (comultdot.315) to (comultoutr.270);
		\end{pgfonlayer}
		\node (comultdef) [right of = comultdot, xshift = 2.8cm] {$:= \; \frac{1}{\sqrt{L}^3}\sum\limits_{n=-\omega}^{+\omega}\sum\limits_{m=-\omega}^{+\omega} \ket{\chi_{n\oplus m}}\;\bra{\chi_n}\otimes\bra{\chi_m}$};
		\node[dot,fill=\Xbwcolour] (counitdot) [right of = comultdef, xshift = 4cm]{};
		\node (counitin) [above of = counitdot] {};
		\begin{pgfonlayer}{background}
		\draw[-] (counitin.90) to (counitdot);
		\end{pgfonlayer}
		\node (counitdef) [right of = counitdot, xshift = 0.75cm] {$:= \;\frac{1}{\sqrt{L}}\ket{\chi_0}$};
	\end{tikzpicture}
	\end{multlined}
\end{equation}

\begin{theorem}\label{thm_GroupAlgebra}
$(\hbox{\begin{tikzpicture} [scale=0.8,transform shape] %% DO NOT CHANGE

\def\deltax{0.3} %% CAN BE CHANGED
\def\deltay{0.5} %% DO NOT CHANGE

%\path[use as bounding box] (-\deltax,-\deltay) rectangle (\deltax,\deltay);

\node (mult_label_outl) at (-\deltax,+\deltay) {};
\node (mult_label_outr) at (+\deltax,+\deltay) {};
\node [dot, fill=\Dcolour] (mult) at (0,0) {};
\node (mult_label_in) at (0,-\deltay) {};
\draw[-] [in=270,out=135] (mult) to (mult_label_outl);
\draw[-] [in=270,out=45] (mult) to (mult_label_outr);
\draw[-] (mult_label_in) to (mult);

%\draw (current bounding box.south west) rectangle (current bounding box.north east);
\end{tikzpicture}}\!,\hbox{\begin{tikzpicture} [scale=0.8,transform shape] %% DO NOT CHANGE

\def\deltax{0.3} %% CAN BE CHANGED
\def\deltay{0.5} %% DO NOT CHANGE

\path[use as bounding box] (-\deltax,-\deltay) rectangle (\deltax,\deltay);

\node [dot, fill=black] (mult) at (0,0) {};
\node (mult_label_in) at (0,-\deltay) {};
\draw[-] (mult_label_in) to (mult);

%\draw (current bounding box.south west) rectangle (current bounding box.north east);
\end{tikzpicture}}\!,\hbox{\begin{tikzpicture} [scale=0.8,transform shape] %% DO NOT CHANGE

\def\deltax{0.3} %% CAN BE CHANGED
\def\deltay{0.5} %% DO NOT CHANGE

%\path[use as bounding box] (-\deltax,-\deltay) rectangle (\deltax,\deltay);

\node (mult_label_inl) at (-\deltax,-\deltay) {};
\node (mult_label_inr) at (+\deltax,-\deltay) {};
\node [dot, fill=\Dcolour] (mult) at (0,0) {};
\node (mult_label_out) at (0,+\deltay) {};

\draw[-] [out=90,in=225](mult_label_inl) to (mult);
\draw[-] [out=90,in=315](mult_label_inr) to (mult);
\draw[-] (mult) to (mult_label_out);

%\draw (current bounding box.south west) rectangle (current bounding box.north east);
\end{tikzpicture}}\!,\hbox{\begin{tikzpicture} [scale=0.8,transform shape] %% DO NOT CHANGE

\def\deltax{0.3} %% CAN BE CHANGED
\def\deltay{0.5} %% DO NOT CHANGE

\path[use as bounding box] (-\deltax,-\deltay) rectangle (\deltax,\deltay);

\node [dot, fill=black] (mult) at (0,0) {};
\node (mult_label_out) at (0,+\deltay) {};
\draw[-] (mult) to (mult_label_out);

%\draw (current bounding box.south west) rectangle (current bounding box.north east);
\end{tikzpicture}}\!)$ is a unital commutative $\dagger$-Frobenius algebra, the \textbf{group algebra} of $\integersMod{2\omega+1}$. Furthermore, it is quasi-special, with normalisation factor $(2\omega + 1)$:
\begin{equation}
	\begin{multlined}
	\begin{tikzpicture}[node distance = 0.5cm]
		\node (comultcenter) {};
		\node[dot,fill=\Xbwcolour] (comultdot) [below of = comultcenter, yshift = 0.25cm] {};
		\node[dot,fill=\Xbwcolour] (comultdotup) [above of = comultcenter, yshift = -0.25cm] {};
		\node (comultin) [below of = comultdot] {};
		\node (comultout) [above of = comultdotup] {}; 
		\begin{pgfonlayer}{background}
		\draw[-] (comultin.270) to (comultdot);
		\draw[-] (comultdotup) to (comultout.90);
		\draw[-,out=135,in=225] (comultdot) to (comultdotup);
		\draw[-,out=45,in=315] (comultdot) to (comultdotup);
		\end{pgfonlayer}
		\node (comultdef) [right of = comultcenter, xshift = 1cm] {$ = \;(2\omega+1)\;\;\;\; $};
		\node (center) [right of = comultdef, xshift = 0.5cm]{};
		\node (in) [below of = center, yshift = -0.25cm] {};
		\node (out) [above of = center, yshift = 0.25cm] {};		
		\begin{pgfonlayer}{background}
		\draw[-] (in.270) to (out.90);
		\end{pgfonlayer}
	\end{tikzpicture}
	\end{multlined}
\end{equation} 
Furthermore, the group algebra for $\integersMod{2\omega + 1}$ copies the (rescaled) position eigenstates:
\begin{equation}
	\begin{multlined}
	\begin{tikzpicture}[node distance = 1cm]
		\node (center) {$=$};
		\node[dot,fill=\Xbwcolour] (lcenter) [left of = center] {};
		\node[point] (lin) [below of = lcenter]{$\delta_x$};
		\node (llabel) [left of = lin,xshift=2mm]{$\sqrt{L}$}; 
		\node (loutl) [above of = lcenter, xshift = -0.5cm] {};
		\node (loutr) [above of = lcenter, xshift = +0.5cm] {};
		\node (rcenter) [right of = center, xshift = 0.5cm]{};
		\node[point] (rinl) [below of = rcenter, xshift = -0.5cm] {$\delta_x$};
		\node (rlabell) [left of = rinl, xshift = 2mm]{$\sqrt{L}$};
		\node (routl) [above of = rinl, yshift = 0.75cm] {};
		\node[point] (rinr) [below of = rcenter, xshift = +0.5cm] {$\delta_x$};
		\node (rlabelr) [right of = rinr, xshift = -4mm]{$\sqrt{L}$};
		\node (routr) [above of = rinr, yshift = 0.75cm] {};
 		\begin{pgfonlayer}{background}
 		\draw[-] (lin) to (lcenter);
 		\draw[-,out=45,in=270] (lcenter.45) to (loutr);
 		\draw[-,out=135,in=270] (lcenter.135) to (loutl);
 		\draw[-] (rinl) to (routl);
 		\draw[-] (rinr) to (routr);
		\end{pgfonlayer}
	\end{tikzpicture}
	\end{multlined}
\end{equation}
As a consequence, we will also refer to it as the \textbf{classical structure for position eigenstates}.
\end{theorem}

We have seen in Theorem \ref{thm_MomentaGenerateTranslations} that momenta generate position-space translation, in the sense that the multiplication $\hbox{\begin{tikzpicture} [scale=0.8,transform shape] %% DO NOT CHANGE

\def\deltax{0.3} %% CAN BE CHANGED
\def\deltay{0.5} %% DO NOT CHANGE

%\path[use as bounding box] (-\deltax,-\deltay) rectangle (\deltax,\deltay);

\node (mult_label_inl) at (-\deltax,-\deltay) {};
\node (mult_label_inr) at (+\deltax,-\deltay) {};
\node [dot, fill=\Zbwcolour] (mult) at (0,0) {};
\node (mult_label_out) at (0,+\deltay) {};

\draw[-] [out=90,in=225](mult_label_inl) to (mult);
\draw[-] [out=90,in=315](mult_label_inr) to (mult);
\draw[-] (mult) to (mult_label_out);

%\draw (current bounding box.south west) rectangle (current bounding box.north east);
\end{tikzpicture}}\!$ and unit $\hbox{\begin{tikzpicture} [scale=0.8,transform shape] %% DO NOT CHANGE

\def\deltax{0.3} %% CAN BE CHANGED
\def\deltay{0.5} %% DO NOT CHANGE

\path[use as bounding box] (-\deltax,-\deltay) rectangle (\deltax,\deltay);

\node [dot, fill=\Zbwcolour] (mult) at (0,-0.25*\deltay) {};
\node (mult_label_out) at (0,+\deltay) {};
\draw[-] (mult) to (mult_label_out);

%\draw (current bounding box.south west) rectangle (current bounding box.north east);
\end{tikzpicture}}\!$ of the classical structure for momentum eigenstates (which, as shown by Theorem \ref{thm_MomentumObservable}, corresponds to the momentum observable) endow the set of position eigenstates with the group structure of position-space translation. The next result (elementary in proof, by the very definition of $\hbox{\begin{tikzpicture} [scale=0.8,transform shape] %% DO NOT CHANGE

\def\deltax{0.3} %% CAN BE CHANGED
\def\deltay{0.5} %% DO NOT CHANGE

%\path[use as bounding box] (-\deltax,-\deltay) rectangle (\deltax,\deltay);

\node (mult_label_inl) at (-\deltax,-\deltay) {};
\node (mult_label_inr) at (+\deltax,-\deltay) {};
\node [dot, fill=\Dcolour] (mult) at (0,0) {};
\node (mult_label_out) at (0,+\deltay) {};

\draw[-] [out=90,in=225](mult_label_inl) to (mult);
\draw[-] [out=90,in=315](mult_label_inr) to (mult);
\draw[-] (mult) to (mult_label_out);

%\draw (current bounding box.south west) rectangle (current bounding box.north east);
\end{tikzpicture}}\!$ and $\hbox{\begin{tikzpicture} [scale=0.8,transform shape] %% DO NOT CHANGE

\def\deltax{0.3} %% CAN BE CHANGED
\def\deltay{0.5} %% DO NOT CHANGE

\path[use as bounding box] (-\deltax,-\deltay) rectangle (\deltax,\deltay);

\node [dot, fill=black] (mult) at (0,0) {};
\node (mult_label_out) at (0,+\deltay) {};
\draw[-] (mult) to (mult_label_out);

%\draw (current bounding box.south west) rectangle (current bounding box.north east);
\end{tikzpicture}}\!$) shows that, dually, positions generate momentum-space translations, in the sense that the multiplication $\hbox{\begin{tikzpicture} [scale=0.8,transform shape] %% DO NOT CHANGE

\def\deltax{0.3} %% CAN BE CHANGED
\def\deltay{0.5} %% DO NOT CHANGE

%\path[use as bounding box] (-\deltax,-\deltay) rectangle (\deltax,\deltay);

\node (mult_label_inl) at (-\deltax,-\deltay) {};
\node (mult_label_inr) at (+\deltax,-\deltay) {};
\node [dot, fill=\Dcolour] (mult) at (0,0) {};
\node (mult_label_out) at (0,+\deltay) {};

\draw[-] [out=90,in=225](mult_label_inl) to (mult);
\draw[-] [out=90,in=315](mult_label_inr) to (mult);
\draw[-] (mult) to (mult_label_out);

%\draw (current bounding box.south west) rectangle (current bounding box.north east);
\end{tikzpicture}}\!$ and unit $\hbox{\begin{tikzpicture} [scale=0.8,transform shape] %% DO NOT CHANGE

\def\deltax{0.3} %% CAN BE CHANGED
\def\deltay{0.5} %% DO NOT CHANGE

\path[use as bounding box] (-\deltax,-\deltay) rectangle (\deltax,\deltay);

\node [dot, fill=black] (mult) at (0,0) {};
\node (mult_label_out) at (0,+\deltay) {};
\draw[-] (mult) to (mult_label_out);

%\draw (current bounding box.south west) rectangle (current bounding box.north east);
\end{tikzpicture}}\!$ of the classical structure for position eigenstates (which Theorem \ref{thm_PositionObservable} below will show to correspond to the position observable) endow the set of momentum eigenstates with the group structure of momentum-space translation.
\begin{theorem}\label{thm_PositionGeneratorsMomenta}
Let $(\hbox{\begin{tikzpicture} [scale=0.8,transform shape] %% DO NOT CHANGE

\def\deltax{0.3} %% CAN BE CHANGED
\def\deltay{0.5} %% DO NOT CHANGE

%\path[use as bounding box] (-\deltax,-\deltay) rectangle (\deltax,\deltay);

\node (mult_label_outl) at (-\deltax,+\deltay) {};
\node (mult_label_outr) at (+\deltax,+\deltay) {};
\node [dot, fill=\Dcolour] (mult) at (0,0) {};
\node (mult_label_in) at (0,-\deltay) {};
\draw[-] [in=270,out=135] (mult) to (mult_label_outl);
\draw[-] [in=270,out=45] (mult) to (mult_label_outr);
\draw[-] (mult_label_in) to (mult);

%\draw (current bounding box.south west) rectangle (current bounding box.north east);
\end{tikzpicture}}\!,\hbox{\begin{tikzpicture} [scale=0.8,transform shape] %% DO NOT CHANGE

\def\deltax{0.3} %% CAN BE CHANGED
\def\deltay{0.5} %% DO NOT CHANGE

\path[use as bounding box] (-\deltax,-\deltay) rectangle (\deltax,\deltay);

\node [dot, fill=black] (mult) at (0,0) {};
\node (mult_label_in) at (0,-\deltay) {};
\draw[-] (mult_label_in) to (mult);

%\draw (current bounding box.south west) rectangle (current bounding box.north east);
\end{tikzpicture}}\!,\hbox{\begin{tikzpicture} [scale=0.8,transform shape] %% DO NOT CHANGE

\def\deltax{0.3} %% CAN BE CHANGED
\def\deltay{0.5} %% DO NOT CHANGE

%\path[use as bounding box] (-\deltax,-\deltay) rectangle (\deltax,\deltay);

\node (mult_label_inl) at (-\deltax,-\deltay) {};
\node (mult_label_inr) at (+\deltax,-\deltay) {};
\node [dot, fill=\Dcolour] (mult) at (0,0) {};
\node (mult_label_out) at (0,+\deltay) {};

\draw[-] [out=90,in=225](mult_label_inl) to (mult);
\draw[-] [out=90,in=315](mult_label_inr) to (mult);
\draw[-] (mult) to (mult_label_out);

%\draw (current bounding box.south west) rectangle (current bounding box.north east);
\end{tikzpicture}}\!,\hbox{\begin{tikzpicture} [scale=0.8,transform shape] %% DO NOT CHANGE

\def\deltax{0.3} %% CAN BE CHANGED
\def\deltay{0.5} %% DO NOT CHANGE

\path[use as bounding box] (-\deltax,-\deltay) rectangle (\deltax,\deltay);

\node [dot, fill=black] (mult) at (0,0) {};
\node (mult_label_out) at (0,+\deltay) {};
\draw[-] (mult) to (mult_label_out);

%\draw (current bounding box.south west) rectangle (current bounding box.north east);
\end{tikzpicture}}\!)$ be the classical structure for position eigenstates. Then the monoid $(\hbox{\begin{tikzpicture} [scale=0.8,transform shape] %% DO NOT CHANGE

\def\deltax{0.3} %% CAN BE CHANGED
\def\deltay{0.5} %% DO NOT CHANGE

%\path[use as bounding box] (-\deltax,-\deltay) rectangle (\deltax,\deltay);

\node (mult_label_inl) at (-\deltax,-\deltay) {};
\node (mult_label_inr) at (+\deltax,-\deltay) {};
\node [dot, fill=\Dcolour] (mult) at (0,0) {};
\node (mult_label_out) at (0,+\deltay) {};

\draw[-] [out=90,in=225](mult_label_inl) to (mult);
\draw[-] [out=90,in=315](mult_label_inr) to (mult);
\draw[-] (mult) to (mult_label_out);

%\draw (current bounding box.south west) rectangle (current bounding box.north east);
\end{tikzpicture}}\!,\hbox{\begin{tikzpicture} [scale=0.8,transform shape] %% DO NOT CHANGE

\def\deltax{0.3} %% CAN BE CHANGED
\def\deltay{0.5} %% DO NOT CHANGE

\path[use as bounding box] (-\deltax,-\deltay) rectangle (\deltax,\deltay);

\node [dot, fill=black] (mult) at (0,0) {};
\node (mult_label_out) at (0,+\deltay) {};
\draw[-] (mult) to (mult_label_out);

%\draw (current bounding box.south west) rectangle (current bounding box.north east);
\end{tikzpicture}}\!)$ endows the set $\suchthat{\ket{\chi_n}}{n \in \integersMod{2\omega+1}}$ of momentum eigenstates with the abelian group structure of momentum-space translation $(\integersMod{2\omega+1},\oplus,0)$.
\begin{equation}
	\begin{multlined}
	\begin{tikzpicture}[node distance = 1cm]
		\node (center) {};
		\node[kpoint, inner sep = 0.35mm] (state) [below of = center] {$\chi_{n \oplus m}$};
		\node (factor) [left of = state] {$\frac{1}{\sqrt{L}}$};
		\node (out) [above of = center] {};
		\node (capdef) [right of = center, xshift = 0.25cm] {$ = $};
		\node[dot,fill=\Xbwcolour] (multcenter) [right of = capdef, xshift = 1cm]{};
		\node[kpoint,inner sep = 0.5mm] (multinl) [below of = multcenter, yshift = 0cm, xshift = -0.75cm] {$\chi_n$};
		\node (factor) [left of = multinl,xshift = 2mm] {$\frac{1}{\sqrt{L}}$};
		\node[kpoint,inner sep = 0.35mm] (multinr) [below of = multcenter, yshift = 0cm, xshift = +0.75cm] {$\chi_m$};
		\node (factor) [right of = multinr, xshift = -2mm] {$\frac{1}{\sqrt{L}}$};
		\node (counit) [above of = multcenter]	{};		
		\begin{pgfonlayer}{background}
		\draw[-] (state) to (out);
		\draw[-,in=225,out=90] (multinl) to (multcenter);
		\draw[-,in=315,out=90] (multinr) to (multcenter);
		\draw[-,out=90,in=270] (multcenter) to (counit);
		\end{pgfonlayer}
	\end{tikzpicture}
	\end{multlined}
\end{equation}
Dually to Theorem \ref{thm_MomentaGenerateTranslations}, we refer to this fact by saying that \textbf{positions generate momentum-space translations}. In particular, the subset of standard momentum eigenstates $\suchthat{\ket{\chi_n}}{n \in \integers}$ is endowed with the abelian group structure of $(\integers,+,0)$.
\end{theorem}

\newpage
Similarly to the momentum observable, the position observable is given by a family $(Q_x)_x$ of orthogonal non-standard projectors on $\Ltwo{\reals/(L\integers)}$ indexed by the set of positions $\suchthat{x}{x \in \reals/(L\integers)}$, where the projectors are 1-dimensional and defined by $Q_x := \frac{L}{2\omega+1}\ket{\delta_x}\bra{\delta_x}$. The following theorem relates this definition to the CQM one in terms of classical structures.

\begin{theorem}\label{thm_PositionObservable}
The position observable can be obtained from the comultiplication $\hbox{\begin{tikzpicture} [scale=0.8,transform shape] %% DO NOT CHANGE

\def\deltax{0.3} %% CAN BE CHANGED
\def\deltay{0.5} %% DO NOT CHANGE

%\path[use as bounding box] (-\deltax,-\deltay) rectangle (\deltax,\deltay);

\node (mult_label_outl) at (-\deltax,+\deltay) {};
\node (mult_label_outr) at (+\deltax,+\deltay) {};
\node [dot, fill=\Dcolour] (mult) at (0,0) {};
\node (mult_label_in) at (0,-\deltay) {};
\draw[-] [in=270,out=135] (mult) to (mult_label_outl);
\draw[-] [in=270,out=45] (mult) to (mult_label_outr);
\draw[-] (mult_label_in) to (mult);

%\draw (current bounding box.south west) rectangle (current bounding box.north east);
\end{tikzpicture}}\!$ of the classical structure for position eigenstates, as follows:
\begin{equation}\label{eqn_PositionObservable}
	\begin{multlined}
	\begin{tikzpicture}[node distance = 1cm]
		\node (center) {$Q_{x}$};
		\node (capdef) [right of = center, xshift = -0.25cm] {$ \simeq $};
		\node[dot,fill=\Xbwcolour] (multcenter) [right of = capdef, xshift = 0cm]{};
		\node (multinl) [above of = multcenter, yshift = 0cm, xshift = -0cm] {};
		\node[copoint,inner sep = 0.1mm] (multinr) [above of = multcenter, yshift = -0.25cm, xshift = +1cm] {$\delta_x$};
		\node (label) [right of = multinr,xshift =-1mm]{$\frac{\sqrt{L}}{\sqrt{2\omega+1}}$};
		\node (counit) [below of = multcenter]	{};		
		\begin{pgfonlayer}{background}
		\draw[-,in=90,out=270] (multinl.90) to (multcenter);
		\draw[-,in=45,out=270] (multinr) to (multcenter);
		\draw[-,out=270,in=90] (multcenter) to (counit);
		\end{pgfonlayer}
	\end{tikzpicture}
	\end{multlined}
\end{equation}
\end{theorem}

\noindent Also, the rescaled counit $\sqrt{L}\hbox{\begin{tikzpicture} [scale=0.8,transform shape] %% DO NOT CHANGE

\def\deltax{0.3} %% CAN BE CHANGED
\def\deltay{0.5} %% DO NOT CHANGE

\path[use as bounding box] (-\deltax,-\deltay) rectangle (\deltax,\deltay);

\node [dot, fill=black] (mult) at (0,0) {};
\node (mult_label_in) at (0,-\deltay) {};
\draw[-] (mult_label_in) to (mult);

%\draw (current bounding box.south west) rectangle (current bounding box.north east);
\end{tikzpicture}}\!$ defines the \textbf{integral operator}: $\sqrt{L}\hbox{\begin{tikzpicture} [scale=0.8,transform shape] %% DO NOT CHANGE

\def\deltax{0.3} %% CAN BE CHANGED
\def\deltay{0.5} %% DO NOT CHANGE

\path[use as bounding box] (-\deltax,-\deltay) rectangle (\deltax,\deltay);

\node [dot, fill=black] (mult) at (0,0) {};
\node (mult_label_in) at (0,-\deltay) {};
\draw[-] (mult_label_in) to (mult);

%\draw (current bounding box.south west) rectangle (current bounding box.north east);
\end{tikzpicture}}\! \ket{\truncate{f}} = \braket{\chi_0}{\truncate{f}} = \int_{\reals/(L\integers)} f(x) dx$.

Finally, in the finite-dimensional case it is known \cite{gogioso2015categorical} that strong complementarity \cite{kissinger2012pictures,coecke2011interacting} corresponds to the Weyl canonical commutation relations, so we expect the classical structures for momentum and position eigenstates to be strongly complementary. This is indeed the case.
\begin{theorem}\label{thm_StrongComplementarity}
The classical structures for momentum and position eigenstates form a strongly complementary pair of unital commutative $\dagger$-Frobenius algebras:
\begin{equation}\label{eqn_StrongComplementarity}
	\begin{multlined}
	\begin{tikzpicture}[node distance = 0.5cm]
		% Bialgebra
		\node (center) {};
		\node[dot, fill=\Zbwcolour] (bl) [below left of = center] {}; 
		\node[dot, fill=\Zbwcolour] (br) [below right of = center] {};
		\node[dot, fill=\Xbwcolour] (tl) [above left of = center] {};
		\node[dot, fill=\Xbwcolour] (tr) [above right of = center] {};
		\node (blin) [below of = bl] {};
		\node (brin) [below of = br] {};
		\node (tlout) [above of = tl] {};
		\node (trout) [above of = tr] {};
		\begin{pgfonlayer}{background}
		\draw[-] (bl) to (tl);
		\draw[-] (bl) to (tr);
		\draw[-] (br) to (tl);
		\draw[-] (br) to (tr);
		\draw[-] (blin) to (bl);
		\draw[-] (brin) to (br);
		\draw[-] (tl) to (tlout);
		\draw[-] (tr) to (trout);
		\end{pgfonlayer}
		\node (equals) [right of = center, xshift = 0.5cm] {$=$};
		\node (center) [right of = equals,xshift = 0.25cm] {};
		\node[dot,fill=\Xbwcolour] (b) [below of = center,yshift=0.25cm] {};
		\node[dot,fill=\Zbwcolour] (t) [above of = center,yshift=-0.25cm] {};
		\node (bl) [below left of = b, yshift = -0.25cm] {};
		\node (br) [below right of = b, yshift = -0.25cm] {};
		\node (tl) [above left of = t, yshift = 0.25cm] {};
		\node (tr) [above right of = t, yshift = 0.25cm] {};
		\begin{pgfonlayer}{background}
		\draw[-] (bl) to (b);
		\draw[-] (br) to (b);
		\draw[-] (t) to (tl);
		\draw[-] (t) to (tr);
		\draw[-] (b) to (t);
		\end{pgfonlayer}
		% Unit coherence
		\node (center) [right of = center,xshift = 1.5cm] {};
		\node[dot,fill=\Xbwcolour] (b) [below of = center,yshift=0.25cm] {};
		\node[dot,fill=\Zbwcolour] (t) [above of = center,yshift=-0.25cm] {};
		\node (tl) [above left of = t, yshift = 0.25cm] {};
		\node (tr) [above right of = t, yshift = 0.25cm] {};
		\begin{pgfonlayer}{background}
		\draw[-] (t) to (tl);
		\draw[-] (t) to (tr);
		\draw[-] (b) to (t);
		\end{pgfonlayer}
		\node (equals) [right of = center, xshift = 0.25cm] {$=$};
		\node (center) [right of = equals,xshift = 0.5cm] {};
		\node[dot,fill=\Xbwcolour] (bl) [below left of = center,yshift=0.1cm] {};
		\node[dot,fill=\Xbwcolour] (br) [below right of = center,yshift=0.1cm] {};
		\node (tl) [above of = bl, yshift = 0.6cm] {};
		\node (tr) [above of = br, yshift = 0.6cm] {};
		\begin{pgfonlayer}{background}
		\draw[-] (bl) to (tl);
		\draw[-] (br) to (tr);
		\end{pgfonlayer}
		% Counit coherence
		\node (center) [right of = center,xshift = 1.5cm] {};
		\node[dot,fill=\Xbwcolour] (b) [below of = center,yshift=0.25cm] {};
		\node[dot,fill=\Zbwcolour] (t) [above of = center,yshift=-0.25cm] {};
		\node (bl) [below left of = b, yshift = -0.25cm] {};
		\node (br) [below right of = b, yshift = -0.25cm] {};
		\begin{pgfonlayer}{background}
		\draw[-] (bl) to (b);
		\draw[-] (br) to (b);
		\draw[-] (b) to (t);
		\end{pgfonlayer}
		\node (equals) [right of = center, xshift = 0.25cm] {$=$};
		\node (center) [right of = equals,xshift = 0.5cm] {};
		\node[dot,fill=\Zbwcolour] (tl) [above left of = center,yshift=-0.1cm] {};
		\node[dot,fill=\Zbwcolour] (tr) [above right of = center,yshift=-0.1cm] {};
		\node (bl) [below of = tl, yshift = -0.6cm] {};
		\node (br) [below of = tr, yshift = -0.6cm] {};
		\begin{pgfonlayer}{background}
		\draw[-] (bl) to (tl);
		\draw[-] (br) to (tr);
		\end{pgfonlayer}
	\end{tikzpicture}
	\end{multlined}
\end{equation}
Equivalently, they are canonically commuting in the sense of the Weyl CCRs (see below for more details):
\begin{equation}\label{eqn_WeylCCRs}
	\begin{multlined}
	\begin{tikzpicture}[node distance = 1cm]
		\node (center) {};
		\node[dot, fill=\Xbwcolour] (b) [below of = center, yshift = 0.5cm]{};
		\node (in) [below of = b] {};
		\node[kpoint] (chi) [below left of = b] {$\chi_n$};
		\node (norm) [left of = chi, xshift = 2mm] {$\frac{1}{\sqrt{L}}$};
		\node[dot, fill=\Zbwcolour] (t) [above of = center, yshift = -0.5cm]{};
		\node (out) [above of = t, yshift = -0.25cm] {};
		\node[point] (delta) [below left of = t] {$\delta_x$};
		\node (norm) [left of = delta, xshift = 2mm] {$\sqrt{L}$};
		\begin{pgfonlayer}{background}
		\draw[-] (in) to (out);
		\draw[-,out=90,in=225] (chi) to (b);
		\draw[-,out=90,in=225] (delta) to (t);
		\end{pgfonlayer}
		\node (equals) [right of = center] {$=$};
		\node (center) [right of = equals, xshift = 1.25cm] {};
		\node[dot, fill=\Zbwcolour] (b) [below of = center, yshift = 0.5cm]{};
		\node (in) [below of = b] {};
		\node[point] (chi) [below left of = b] {$\delta_x$};
		\node (norm) [left of = chi, xshift = 2mm] {$\sqrt{L}$};
		\node[dot, fill=\Xbwcolour] (t) [above of = center, yshift = -0.5cm]{};
		\node (out) [above of = t, yshift = -0.25cm] {};
		\node[kpoint] (delta) [below left of = t] {$\chi_n$};
		\node (norm) [left of = delta, xshift = 2mm] {$\frac{1}{\sqrt{L}}$};
		\begin{pgfonlayer}{background}
		\draw[-] (in) to (out);
		\draw[-,out=90,in=225] (chi) to (b);
		\draw[-,out=90,in=225] (delta) to (t);
		\end{pgfonlayer}
		\node[scalar] (factor) [right of = center, xshift = -0.1cm]{$\chi_n(x)$};
	\end{tikzpicture}
	\end{multlined}
\end{equation}
\end{theorem}
\noindent Traditionally, the Weyl CCRs take the following form:
\begin{equation}%\nonumber
\exp\big[i (2\pi/L) x \,\frac{\textbf{p}}{\hbar}\big]\cdot\exp\big[i (2\pi/L) \frac{p}{\hbar} \,\textbf{x}\big] = e^{i (2 \pi / L) x\,\frac{p}{\hbar}}\exp\big[i (2\pi/L) \frac{p}{\hbar}\, \textbf{x}\big]\cdot\exp\big[i (2\pi/L) x \,\frac{\textbf{p}}{\hbar}\big], 
\end{equation}
However, the traditional formulation requires the position and momentum observables to be given in the self-adjoint form of infinitesimal generators. When observables are given by complete families of orthogonal projectors, a different form is required: contrary to the differential formulation of the CCRs in terms of commutators, the Weyl CCRs are easily adapted to this alternative notion of observable:  
\begin{equation}\nonumber
\Big[ \sum_{m} e^{i (\frac{2 \pi}{L}) m x} \ket{\chi_m}\bra{\chi_m}\Big] \cdot \Big[ \int e^{i (\frac{2 \pi}{L}) n y} \ket{\delta_y}\bra{\delta_y}dy\Big] = e^{-i(\frac{2 \pi}{L})n x } \Big[ \int e^{i (\frac{2 \pi}{L}) n y} \ket{\delta_y}\bra{\delta_y}dy\Big] \cdot \Big[ \sum_{m} e^{i (\frac{2 \pi}{L}) m x} \ket{\chi_m}\bra{\chi_m}\Big]
\end{equation}
The Weyl CCRs in this new form are exactly those given by Equation \ref{eqn_WeylCCRs} (once we write them down linearly, in semi-rigorous notation).

In this section, we have seen how the traditional ingredients of quantum mechanics for wavefunctions on $\Ltwo{\reals/(L \integers)}$ find a natural place in the structures of infinite-dimensional categorical quantum mechanics. The methods presented here can be extended to the case of wavefunctions on spaces with a compact or discrete abelian group of translations (such as tori or lattices). A detailed treatment is left to future work.

\newpage
\section{Conclusions}
\vspace{-0.1cm}
Using tools from non-standard analysis, we have constructed a new category $\starHilbCategory$. This category extends Hilbert spaces and bounded linear maps in a suitable sense, and contains all the structures required to do categorical quantum mechanics in separable Hilbert spaces (such as unital commutative $\dagger$-Frobenius algebras and dagger compact structure). As a sample application, we have covered the textbook case of a 1-dimensional wavefunction with periodic boundary condition: the same principles that govern observables in finite dimensions also allow us to obtain the usual (unbounded) position and momentum observables. The work presented here is only the beginning of what promises to be a long and exciting venture, but it already provides evidence that the same abstract methods that have made the framework so successful throughout the last decade can be extended to infinite dimensions, with little modification. We are confident that further developments will help close the gap between the categorical framework and the formalism needs of the practising quantum physicist.

Our non-standard extension of categorical quantum mechanics is successful, but it comes at a cost: the very objects that contribute to the extension -- unbounded operators, Dirac deltas, plane-waves and infinitesimal probabilities -- are in direct contrast with the purportedly finite / continuous nature of concrete physical experience. To this point, our approach is mainly a pragmatic one, rather than a foundational one: we have chosen to meet the flexibility required by everyday practice of quantum mechanics with the rigour of categorical constructions. To those who believe that plane-waves and Dirac deltas have physical significance, $\starHilbCategory$ may represent physical reality better than $\HilbCategory$; to those who don't, it provides a space where common mathematical \inlineQuote{tricks} turn out to be well-defined. There certainly is a deeply interesting story behind the divergences of unbounded operators and Dirac deltas, but we leave that for another day. Today, we chose to extend the reach of categorical and diagrammatic methods to the (separable) infinite-dimensional world that most quantum physics takes place in. We hope you enjoyed the ride.

\vspace{-0.1cm}
%\newpage
%\nocite{*}
%\bibliography{biblio}

\newpage
\appendix

\section{Proofs}

\subsection{Proofs from Section \ref{section_StarHilbCategorydef}}

\paragraph{Proof of associativity for the tensor product}
In order to show that the tensor product defined by Equation (\ref{eqn_TensorObjects}) is associative, all we have to show is that the following two bijections $\{1,...,\kappa\}\times\{1,...,\nu\}\times\{1,...,\lambda\} \rightarrow \{1,...,\kappa \nu \lambda\}$ coincide:
\begin{align}
(n,m,l) \mapsto \varsigma_{\kappa\nu,\lambda}\big(\varsigma_{\kappa,\nu}(n,m),l\big) \nonumber \\
(n,m,l) \mapsto \varsigma_{\kappa,\nu\lambda}\big(n,\varsigma_{\nu,\lambda}(m,l)\big)
\end{align}
This is a simple matter of non-standard algebra:
\begin{align}
 \varsigma_{\kappa\nu,\lambda}\big(\varsigma_{\kappa,\nu}(n,m),l\big) &=  \Big(\big((n-1)\nu + m\big)-1\Big)\lambda+ l \,= \nonumber \\ 
 &= (n-1)\nu\lambda + \big((m-1)\lambda + l\big) = \varsigma_{\kappa,\nu\lambda}\big(n,\varsigma_{\nu,\lambda}(m,l)\big)
\end{align}
\hfill$\square$
\vspace{-0.5cm}

\subsection{Proofs from Section \ref{section_StandardBoundedMaps}}

\paragraph{Proof of Lemma \ref{thm_Continuity}.}
\textit{(i) implies (ii)}: let $\zeta := \braket{\xi_\kappa}{\xi_\kappa}$ be an infinitesimal; then we have $\bra{\xi_\kappa} \truncate{F}^\dagger \truncate{F} \ket{\xi_\kappa} \leq \zeta ||\truncate{F}||_{op}$, which is infinitesimal since $||\truncate{F}||_{op}$ is finite. \textit{(ii) implies (i)}: if $||\truncate{F}||_{op}$ is infinite, then for some $\ket{\psi_\kappa}$ of unit norm we have $\bra{\psi_\kappa}\truncate{F}^\dagger \truncate{F} \ket{\psi_\kappa} = \theta$ infinite; but then $\bra{\psi_\kappa}\frac{1}{\sqrt{\theta}}\truncate{F}^\dagger \truncate{F} \frac{1}{\sqrt{\theta}}\ket{\psi_\kappa} = 1$ is not infinitesimal, with $\frac{1}{\sqrt{\theta}}\ket{\psi_\kappa}$ infinitesimal. \textit{(ii) equivalent to (iii)}: by linearity of $\truncate{F}$.
\hfill$\square$

\paragraph{Proof of Lemma \ref{thm_InfinitesimalOpEquiv}}
Bi-linearity of composition and tensor product, together with the triangle inequality, imply that the only statements we need to prove are the following:
\begin{align}
||\truncate{G} \cdot \xi_\kappa||_{op} &\text{ infinitesimal whenever } \truncate{G} \text{ continuous and } \xi_\kappa \text{ infinitesimal;} \nonumber \\
||\zeta_\kappa \cdot \truncate{F}||_{op} &\text{ infinitesimal whenever } \truncate{F} \text{ continuous and } ||\zeta_\kappa||_{op} \text{ infinitesimal;} \nonumber \\
||\truncate{G} \otimes \xi_\kappa||_{op} &\text{ infinitesimal whenever } \truncate{G} \text{ continuous and } ||\xi_\kappa||_{op} \text{ infinitesimal.}
\end{align}
The first statement follows from the fact that $||\truncate{G} \xi_\kappa||_{op} \leq ||\truncate{G}||_{op} ||\xi_\kappa||_{op}$, which is infinitesimal because $||\truncate{G}||_{op}$ is finite. The second statement goes similarly. The third statement is slightly trickier. Let $\ket{\psi_\kappa}$ be unit norm, and write $\ket{\psi_\kappa} = \sum_{n} \ket{\phi_\kappa^{(n)}} \ket{e_{n}}$ (where $(\ket{e_{n}})_{n}$ is the chosen orthonormal basis for the domain of $\xi_\kappa$). Then we have the following
\begin{equation}
\bra{\psi_\kappa} (\truncate{G} \otimes \xi_\kappa)^\dagger (\truncate{G} \otimes \xi_\kappa) \ket{\psi_\kappa} \leq \sum_{n'} \sum_{n} \braket{\phi_\kappa^{(n)}}{\phi_\kappa^{(n)}} ||\truncate{G}||_{op} |(\xi_\kappa)_{n'n}|^2 \leq ||\truncate{G}||_{op} ||\xi_\kappa||_{op},
\end{equation}
where the last product is infinitesimal because $||\truncate{G}||_{op}$ is finite.
\hfill$\square$

\paragraph{Existence and uniqueness of definition of $\stdpart{\truncate{F}}$}
By definition, if $\truncate{F}$ is near-standard, at least one standard bounded linear map $f'$ exists such that $\truncate{F} \sim \truncate{f'}$. Now take two such standard bounded linear maps $f'$ and $f''$: by transitivity we get that $f' \sim f''$, i.e. that $||\truncate{f'}-\truncate{f''}||_{op}$ is infinitesimal; define $g := f'-f''$, standard bounded linear map. By transfer theorem (both directions), $\sqrt{\bra{\psi}g^\dagger g\ket{\psi}}$ is bounded above (by a standard constant $c \in \reals^+$, for all standard $\ket{\psi}$ satisfying $\braket{\psi}{\psi}=1$), if and only if $\sqrt{\bra{\psi}g^\dagger g\ket{\psi}}$ is also bounded above (by the same standard constant $c$, for all internal $\ket{\psi}$ such that $\braket{\psi}{\psi}=1$). Because $g$ is standard and bounded, $\sqrt{\bra{\psi}g^\dagger g\ket{\psi}}$ and $\sqrt{\bra{\psi}\truncate{g}^\dagger \truncate{g}\ket{\psi}}$ are infinitesimally close: as a consequence, if $||\truncate{f'}-\truncate{f''}||_{op}$ is infinitesimal, then it is bounded above by all standard reals $c>0$, and hence by transfer theorem so is $||f'-f''||_{op}$. This proves that $||f'-f''||_{op}=0$, and we conclude that $f'=f''$.
\hfill$\square$

\paragraph{Choice of orthonormal bases for $\vartruncate{\emptyArg}_\omega$}
Up to equivalence of categories, we can consider $\sHilbCategory$ as having objects given by all finite (possibly empty) tensor products of the following basic objects: the finite-dimensional Hilbert spaces $\complexs^p$ for all primes $p$, and the separable space $\ltwo{\naturals^+}$. Choose any orthonormal basis for each of the basic objects; denote them by $\ket{e^{(p)}_n}_{n=1}^p$ and $\ket{e^{(\infty)}_n}_{n=1}^{\infty}$. On basic objects, define $\vartruncate{\complexs^p}_\omega := (\complexs^p,\ket{e^{(p)}_n}_{n=1}^p)$ and $\vartruncate{\ltwo{\naturals^+}}_\omega := (\ltwo{\naturals^+},\ket{e^{(\infty)}_n}_{n=1}^\omega)$. Extend the definition to finite tensor products by using the tensor product of $\starHilbCategory$ (or, equivalently, by using the bijection $\varsigma$ from Equation (\ref{eqn_varsigma}) to explicitly construct a basis). 
\hfill$\square$

\paragraph{Proof that $f \mapsto \truncate{f}$ is an injection}
Let $f,g$ be standard bounded linear maps, defined by matrices $(a_{nm})_{n,m\in \naturals^+}$ and $(b_{nm})_{n,m \in \naturals^+}$ respectively. The matrices can be extended by transfer theorem to non-standard indices, and $\truncate{f}$ and $\truncate{g}$ have matrices $(a_{nm})_{\varsigma(n,m)=1}^{\kappa\nu}$ and $(b_{nm})_{\varsigma(n,m)=1}^{\kappa\nu}$. If $\truncate{f}=\truncate{g}$, then we have that $(a_{nm})_{\varsigma(n,m)=1}^{\kappa\nu}=(b_{nm})_{\varsigma(n,m)=1}^{\kappa\nu}$ as matrices, and in particular $a_{nm}=b_{nm}$ for all $n,m \in \naturals^+$, proving that $f=g$ in the first place. 

\paragraph{Weak functoriality of $\vartruncate{\emptyArg}_\omega$}
We begin by covering weak functoriality of $\vartruncate{\emptyArg}_\omega$, as it makes an interesting point by itself. Note that $\vartruncate{g}_\omega \cdot \vartruncate{f}_\omega = P_\SpaceG \circ g \circ P_\SpaceH \circ f \circ P_\SpaceK$, and that $\vartruncate{g\cdot f} = P_\SpaceG \circ g \circ f \circ P_\SpaceK$. In the infinite-dimensional case, if $f,g$ are standard bounded linear maps, then the standard series $a_{ln} := \sum_{m=0}^\infty g_{lm}f_{mn}$ converges for all fixed $l,n$, and the non-standard complex number $\sum_{m=0}^\kappa g_{lm}f_{mn}$ is infinitesimally close to the standard complex number $a_{ln}$. Hence $g \circ P_\SpaceH \circ g \sim g \circ f$, when seen as internal morphisms of non-standard Hilbert spaces. In the finite-dimensional case, there is no issue of truncation, and $\vartruncate{\emptyArg}_\omega$ is strictly functorial.
\hfill$\square$

\paragraph{Proof of Theorem \ref{thm_SeparableInStarHilb}}
\begin{enumerate}
\item[(i)] The map $\stdpart{\emptyArg}$ is well-defined and monoidally functorial by Lemma \ref{thm_InfinitesimalOpEquiv}. It is full by the proof of existence/uniqueness given above, and surjective on objects by construction of $\starHilbCategory$ and $\sHilbCategory$.
\item[(ii)] The map $\vartruncate{\emptyArg}_{\omega}$ is weakly functorial by the argument given at the beginning of this proof (strictly functorial when restricted to $\fdHilbCategory$), and monoidally so by Lemma \ref{thm_InfinitesimalOpEquiv} and the choice of orthonormal bases presented above. Faithfulness was proven above (by showing that $f \mapsto \truncate{f}$ is an injection), and essential surjectivity follows from point (v) below.
\item[(iii)] We know from above that $\vartruncate{\emptyArg}_\omega$ is faithful, i.e. that $f \mapsto \truncate{f}$ is an injection. If $f$ is a standard bounded linear map, then the morphism $\stdpart{\vartruncate{f}_\omega}$ of $\sHilbCategory$ is the unique standard bounded linear map which is infinitesimally close (in operator norm) to $\truncate{f}$, i.e. it is $f$ itself.
\item[(iv)] By definition of the two functors.
\item[(v)] By (iii), one such standard unitary $\truncate{u}_\SpaceH$ exists, namely by taking $u := \id{\SpaceH}$. Uniqueness follows because any such unitary must be infinitesimally close to the standard unitary $\truncate{u}_\SpaceH$ define above, and at most one such standard linear map exists.
\item[(vi)] The morphism $\truncate{u}_{\SpaceG}^\dagger \vartruncate{\stdpart{\truncate{F}}}_\omega \truncate{u}_\SpaceH$ is infinitesimally close to its image under $\stdpart{\emptyArg}$, which is $\stdpart{\truncate{F}}$ by points (iii) and (v) above. Similarly, $\truncate{F}$ is infinitesimally close to its image under $\stdpart{\emptyArg}$, which is also $\stdpart{\truncate{F}}$. We conclude by transitivity/symmetry of $\sim$. 
\end{enumerate}
\hfill$\square$

\subsection{Proofs from Section \ref{section_ClassicalStructures}}

\paragraph{Proof of Theorem \ref{thm_ClassicalStructures}}
Associativity and Frobenius laws hold with strict equalities (not up to $\sim$, despite involving composition of standard bounded linear maps), exactly as shown in \cite{abramsky2012hstaralgebras}. Commutativity also holds with strict equality. The only things left to check are a Unit law and the Speciality law.
\begin{equation}
	\begin{multlined}
	\begin{tikzpicture}[node distance = 0.5cm]
		\node[dot,fill=\Zbwcolour] (comultdot) {};
		\node (comultin) [below of = comultdot] {};
		\node (comultoutl) [above of = comultdot, xshift = -0.5cm] {}; 
		\node[dot,fill=\Zbwcolour] (comultoutr) [above of = comultdot, xshift = +0.5cm] {};
		\begin{pgfonlayer}{background}
		\draw[-] (comultin.270) to (comultdot);
		\draw[-,out=135,in=270] (comultdot.135) to (comultoutl.90);
		\draw[-,out=45,in=270] (comultdot.45) to (comultoutr);
		\end{pgfonlayer}
		\node (comultdef) [right of = comultdot, xshift = 3.25cm] {$=(\id{\SpaceH} \otimes \sum\limits_{m=1}^\kappa\bra{f_m})\cdot \sum\limits_{n=1}^\kappa \ket{f_n}\otimes\ket{f_n}\;\bra{f_n} \sim$};
		\node (comultdef2) [below of = comultdef,yshift = -0.75cm, xshift = 0.3cm] {$\sim \sum\limits_{m=1}^\kappa\sum\limits_{n=1}^\kappa \ket{f_n}\braket{f_m}{f_n}\bra{f_n} = \sum\limits_{n=1}^\kappa \ket{f_n}\bra{f_n} \sim $};
		\node (center) [right of = comultdef2, xshift = 3cm]{};
		\node (in) [below of = center] {};
		\node (out) [above of = center] {};		
		\begin{pgfonlayer}{background}
		\draw[-] (in.270) to (out.90);
		\end{pgfonlayer}
	\end{tikzpicture}
	\end{multlined}
\end{equation}
\vspace{-0.2mm}
\begin{equation}
	\begin{multlined}
	\hspace{0.25cm}
	\begin{tikzpicture}[node distance = 0.5cm]
		\node (comultcenter) {};
		\node[dot,fill=\Zbwcolour] (comultdot) [below of = comultcenter, yshift = 0.25cm] {};
		\node[dot,fill=\Zbwcolour] (comultdotup) [above of = comultcenter, yshift = -0.25cm] {};
		\node (comultin) [below of = comultdot] {};
		\node (comultout) [above of = comultdotup] {}; 
		\begin{pgfonlayer}{background}
		\draw[-] (comultin.270) to (comultdot);
		\draw[-] (comultdotup) to (comultout.90);
		\draw[-,out=135,in=225] (comultdot) to (comultdotup);
		\draw[-,out=45,in=315] (comultdot) to (comultdotup);
		\end{pgfonlayer}
		\node (comultdef) [right of = comultcenter, xshift = 3.25cm] {$=(\sum\limits_{m=1}^\kappa\ket{f_m}\;\bra{f_m}\otimes\bra{f_m})\cdot (\sum\limits_{n=1}^\kappa \ket{f_n}\ket{f_n}\bra{f_n}) \sim$};
		\node (comultdef2) [below of = comultdef,yshift = -0.75cm, xshift = 0.1cm] {$\sim \sum\limits_{m=1}^\kappa\sum\limits_{n=1}^\kappa \ket{f_m}\braket{f_m}{f_n}^2\bra{f_n} = \sum\limits_{n=1}^\kappa \ket{f_n}\bra{f_n} \sim $};
		\node (center) [right of = comultdef2, xshift = 3cm]{};
		\node (in) [below of = center] {};
		\node (out) [above of = center] {};		
		\begin{pgfonlayer}{background}
		\draw[-] (in.270) to (out.90);
		\end{pgfonlayer}
	\end{tikzpicture}
	\end{multlined}
\end{equation}
Finally, if $\ket{f_n}_n$ is the chosen orthonormal basis $\ket{e_n}_n$ for $\SpaceH$, then the $\sim$ in the previous equations are in fact $=$, and the classical structure is a strictly unital, strictly special commutative $\dagger$-Frobenius algebra\footnote{\label{footnote_SimIntoEq}The $\sim$ become $=$ because the identity takes the exact form $\id{\SpaceH} = \sum_{n=1}^\kappa \ket{f_n}\bra{f_n}$, rather than the approximate form $\id{\SpaceH} \sim \sum_{n=1}^\kappa \ket{f_n}\bra{f_n}$, when $\ket{f_n}_{n=1}^\kappa$ is the chosen ort'l basis.}.
\hfill$\square$

\vspace{-0.4cm}

\paragraph{Proof of Theorem \ref{thm_CompactClosed}}
Weak yanking equations follow from the Frobenius law and weak Unit laws of any classical structure in $\starHilbCategory$. When the classical structure is that of the chosen orthonormal basis, the strict Unit laws result in strict yanking equations, yielding legitimate cups and caps (again because of the exact resolution of the identity into $\id{\SpaceH} = \sum_{n=1}^\kappa \ket{f_n}\bra{f_n}$).  
\hfill$\square$

\vspace{-0.2cm}

\subsection{Proofs from Section \ref{section_WavefunctionsPeriodic}}

\paragraph{Proof of Theorem \ref{thm_PositionEigenstates}}
The proof that the state $\ket{\delta_{x_0}}$ satisfies $\braket{\delta_{x_0}}{f} \simeq f(0)$ for all standard smooth $f \in \Ltwo{\reals/L\integers}$ hinges on the transfer theorem, together with the following standard result from Fourier theory:
\begin{equation}
\frac{1}{\sqrt{L}}\sum_{n=-N}^{N} e^{-i(2\pi/L)x_0n}\frac{1}{\sqrt{L}}\braket{\chi_n}{f} = \frac{1}{L}\int_{\reals/L\integers} \Big(\sum_{n=-N}^{N} e^{i(2\pi/L)(x-x_0)n}\Big) f(x) dx\; \stackrel{N \rightarrow \infty}{\longrightarrow}\; f(x_0).
\end{equation}
To show orthogonality, we consider $\ket{f} := \sum_{m=-M}^{M}\big(\frac{1}{L}e^{i(2\pi/L)x_1 m}\big)\ket{\chi_m}$ for some $x_1 \neq x_0$. This isn't a smooth function, so the result we just obtained cannot be applied to it; however, a similar reasoning (this time with two limits and two applications of the transfer theorem) yields the desired result:
\begin{align}
&\frac{1}{\sqrt{L}}\sum_{n=-N}^{N}\frac{1}{\sqrt{L}}\sum_{m=-M}^{M} \big(e^{-i(2\pi/L)x_0n}\frac{1}{\sqrt{L}}\big)\braket{\chi_n}{\chi_m}\big(e^{i(2\pi/L)x_1 m} \frac{1}{\sqrt{L}}\big) =\nonumber \\
= &\frac{1}{L^2}\int_{\reals/L\integers} \Big(\sum_{n=-N}^{N}\sum_{m=-M}^{M} e^{i(2\pi/L)\big((x-x_0)n+(x-x_1)m\big)}\Big) dx\; \stackrel{N,M \rightarrow \infty}{\longrightarrow}\; 0.
\end{align}
Finally, the position eigenstates are clearly unbiased for the momentum eigentstates: by the first part of this proof, any given position eigenstate $\ket{\delta_{x_0}}$ satisfies $|\braket{\delta_{x_0}}{\chi_n}|^2 \simeq 1$ for all momentum eigenstates $\ket{\chi_n}$, independently of $n$.
\hfill$\square$

\paragraph{Proof of Theorem \ref{thm_MomentaGenerateTranslations}}
Using the definition of the classical structure for momentum eigenstates, together with the definition of the position eigenstates, we obtain the desired equalities:
\begin{align}
\hbox{\begin{tikzpicture} [scale=0.8,transform shape] %% DO NOT CHANGE

\def\deltax{0.3} %% CAN BE CHANGED
\def\deltay{0.5} %% DO NOT CHANGE

%\path[use as bounding box] (-\deltax,-\deltay) rectangle (\deltax,\deltay);

\node (mult_label_inl) at (-\deltax,-\deltay) {};
\node (mult_label_inr) at (+\deltax,-\deltay) {};
\node [dot, fill=\Zbwcolour] (mult) at (0,0) {};
\node (mult_label_out) at (0,+\deltay) {};

\draw[-] [out=90,in=225](mult_label_inl) to (mult);
\draw[-] [out=90,in=315](mult_label_inr) to (mult);
\draw[-] (mult) to (mult_label_out);

%\draw (current bounding box.south west) rectangle (current bounding box.north east);
\end{tikzpicture}}\! \cdot (\sqrt{L}\ket{\delta_x}\sqrt{L}\ket{\delta_y}) &= \Big[\frac{1}{\sqrt{L}^3}\sum_{n=-\omega}^{+\omega} \ket{\chi_n}\bra{\chi_n}\bra{\chi_n} \Big]\sqrt{L}\ket{\delta_x}\sqrt{L}\ket{\delta_y} = \Big[\frac{1}{\sqrt{L}^2}\sum_{n=-\omega}^{+\omega} \chi_{n}(x)^\star \ket{\chi_n}\bra{\chi_n} \Big] \sqrt{L}\ket{\delta_y} = \nonumber \\
&= \frac{1}{\sqrt{L}}\sum_{n=-\omega}^{+\omega} \chi_{n}(x)^\star \chi_{n}(y)^\star \ket{\chi_n} = \frac{1}{\sqrt{L}}\sum_{n=-\omega}^{+\omega} \chi_{n}(x \oplus y)^\star \ket{\chi_n} = \sqrt{L}\ket{\delta_{x \oplus y}}.
\end{align}
\hfill$\square$

% \paragraph{Proof of Theorem \ref{thm_MomentumObservable}}
% A straightforward check.
% \hfill$\square$
\vspace{-0.6cm}

\paragraph{Proof of Theorem \ref{thm_GroupAlgebra}}
Commutative, Associative and Unit laws can be proven on the monoid using the corresponding laws for $(\oplus,0)$. Frobenius law follows from the following re-indexing, with $k' := k\oplus n$:
\begin{align}
(\hbox{\begin{tikzpicture} [scale=0.8,transform shape] %% DO NOT CHANGE

\def\deltax{0.3} %% CAN BE CHANGED
\def\deltay{0.5} %% DO NOT CHANGE

%\path[use as bounding box] (-\deltax,-\deltay) rectangle (\deltax,\deltay);

\node (mult_label_inl) at (-\deltax,-\deltay) {};
\node (mult_label_inr) at (+\deltax,-\deltay) {};
\node [dot, fill=\Dcolour] (mult) at (0,0) {};
\node (mult_label_out) at (0,+\deltay) {};

\draw[-] [out=90,in=225](mult_label_inl) to (mult);
\draw[-] [out=90,in=315](mult_label_inr) to (mult);
\draw[-] (mult) to (mult_label_out);

%\draw (current bounding box.south west) rectangle (current bounding box.north east);
\end{tikzpicture}}\! \otimes \id{})\cdot(\id{} \otimes \hbox{\begin{tikzpicture} [scale=0.8,transform shape] %% DO NOT CHANGE

\def\deltax{0.3} %% CAN BE CHANGED
\def\deltay{0.5} %% DO NOT CHANGE

%\path[use as bounding box] (-\deltax,-\deltay) rectangle (\deltax,\deltay);

\node (mult_label_outl) at (-\deltax,+\deltay) {};
\node (mult_label_outr) at (+\deltax,+\deltay) {};
\node [dot, fill=\Dcolour] (mult) at (0,0) {};
\node (mult_label_in) at (0,-\deltay) {};
\draw[-] [in=270,out=135] (mult) to (mult_label_outl);
\draw[-] [in=270,out=45] (mult) to (mult_label_outr);
\draw[-] (mult_label_in) to (mult);

%\draw (current bounding box.south west) rectangle (current bounding box.north east);
\end{tikzpicture}}\!) &= \frac{1}{\sqrt{L}^4}\sum_{n=-\omega}^{+\omega} \sum_{m=-\omega}^{+\omega} \Big[\;\,\sum_{k\oplus h = m} \ket{\chi_{n\oplus k}}\otimes\ket{\chi_{h}}\;\bra{\chi_n}\otimes\bra{\chi_m}\Big] = \nonumber \\
&= \frac{1}{\sqrt{L}^4}\sum_{n=-\omega}^{+\omega} \sum_{m=-\omega}^{+\omega}\Big[\sum_{k'\oplus h = n \oplus m} \ket{\chi_{k'}}\otimes\ket{\chi_{h}}\;\bra{\chi_n}\otimes\bra{\chi_m}\Big] = \hbox{\begin{tikzpicture} [scale=0.8,transform shape] %% DO NOT CHANGE

\def\deltax{0.3} %% CAN BE CHANGED
\def\deltay{0.5} %% DO NOT CHANGE

%\path[use as bounding box] (-\deltax,-\deltay) rectangle (\deltax,\deltay);

\node (mult_label_outl) at (-\deltax,+\deltay) {};
\node (mult_label_outr) at (+\deltax,+\deltay) {};
\node [dot, fill=\Dcolour] (mult) at (0,0) {};
\node (mult_label_in) at (0,-\deltay) {};
\draw[-] [in=270,out=135] (mult) to (mult_label_outl);
\draw[-] [in=270,out=45] (mult) to (mult_label_outr);
\draw[-] (mult_label_in) to (mult);

%\draw (current bounding box.south west) rectangle (current bounding box.north east);
\end{tikzpicture}}\! \cdot \hbox{\begin{tikzpicture} [scale=0.8,transform shape] %% DO NOT CHANGE

\def\deltax{0.3} %% CAN BE CHANGED
\def\deltay{0.5} %% DO NOT CHANGE

%\path[use as bounding box] (-\deltax,-\deltay) rectangle (\deltax,\deltay);

\node (mult_label_inl) at (-\deltax,-\deltay) {};
\node (mult_label_inr) at (+\deltax,-\deltay) {};
\node [dot, fill=\Dcolour] (mult) at (0,0) {};
\node (mult_label_out) at (0,+\deltay) {};

\draw[-] [out=90,in=225](mult_label_inl) to (mult);
\draw[-] [out=90,in=315](mult_label_inr) to (mult);
\draw[-] (mult) to (mult_label_out);

%\draw (current bounding box.south west) rectangle (current bounding box.north east);
\end{tikzpicture}}\!
\end{align}
Finally, the algebra is quasi-special, with normalisation factor $(2 \omega + 1)$:
\begin{align}
\hbox{\begin{tikzpicture} [scale=0.8,transform shape] %% DO NOT CHANGE

\def\deltax{0.3} %% CAN BE CHANGED
\def\deltay{0.5} %% DO NOT CHANGE

%\path[use as bounding box] (-\deltax,-\deltay) rectangle (\deltax,\deltay);

\node (mult_label_inl) at (-\deltax,-\deltay) {};
\node (mult_label_inr) at (+\deltax,-\deltay) {};
\node [dot, fill=\Dcolour] (mult) at (0,0) {};
\node (mult_label_out) at (0,+\deltay) {};

\draw[-] [out=90,in=225](mult_label_inl) to (mult);
\draw[-] [out=90,in=315](mult_label_inr) to (mult);
\draw[-] (mult) to (mult_label_out);

%\draw (current bounding box.south west) rectangle (current bounding box.north east);
\end{tikzpicture}}\! \cdot \hbox{\begin{tikzpicture} [scale=0.8,transform shape] %% DO NOT CHANGE

\def\deltax{0.3} %% CAN BE CHANGED
\def\deltay{0.5} %% DO NOT CHANGE

%\path[use as bounding box] (-\deltax,-\deltay) rectangle (\deltax,\deltay);

\node (mult_label_outl) at (-\deltax,+\deltay) {};
\node (mult_label_outr) at (+\deltax,+\deltay) {};
\node [dot, fill=\Dcolour] (mult) at (0,0) {};
\node (mult_label_in) at (0,-\deltay) {};
\draw[-] [in=270,out=135] (mult) to (mult_label_outl);
\draw[-] [in=270,out=45] (mult) to (mult_label_outr);
\draw[-] (mult_label_in) to (mult);

%\draw (current bounding box.south west) rectangle (current bounding box.north east);
\end{tikzpicture}}\! &= \frac{1}{\sqrt{L}^2}\sum_{n=-\omega}^{+\omega}  \ket{\chi_n} \Big( \sum_{k\oplus h = n} \frac{1}{\sqrt{L}^2}\braket{\chi_k}{\chi_k} \frac{1}{\sqrt{L}^2}\braket{\chi_h}{\chi_h} \Big) \bra{\chi_n} = \nonumber \\
&= \frac{1}{\sqrt{L}^2}\sum_{n=-\omega}^{+\omega}  \Big( 2\omega + 1\Big) \bra{\chi_n} = (2 \omega + 1) \id{}.
\end{align}
The fact that position eigenstates are copied is a straightforward check, with a re-indexing $n' := n \ominus k$ in the second-to-last step:
\begin{align}
\hbox{\begin{tikzpicture} [scale=0.8,transform shape] %% DO NOT CHANGE

\def\deltax{0.3} %% CAN BE CHANGED
\def\deltay{0.5} %% DO NOT CHANGE

%\path[use as bounding box] (-\deltax,-\deltay) rectangle (\deltax,\deltay);

\node (mult_label_outl) at (-\deltax,+\deltay) {};
\node (mult_label_outr) at (+\deltax,+\deltay) {};
\node [dot, fill=\Dcolour] (mult) at (0,0) {};
\node (mult_label_in) at (0,-\deltay) {};
\draw[-] [in=270,out=135] (mult) to (mult_label_outl);
\draw[-] [in=270,out=45] (mult) to (mult_label_outr);
\draw[-] (mult_label_in) to (mult);

%\draw (current bounding box.south west) rectangle (current bounding box.north east);
\end{tikzpicture}}\! \cdot \sqrt{L} \ket{\delta_x} &= \frac{1}{\sqrt{L}^2}\sum_{n=-\omega}^{+\omega} \sum_{k=-\omega}^{+\omega} \ket{\chi_k}\otimes \ket{\chi_{n\ominus_k}}\; \braket{\chi_n}{\delta_x} = \nonumber \\
&= \frac{1}{\sqrt{L}^2}\sum_{n=-\omega}^{+\omega} \sum_{k=-\omega}^{+\omega} \ket{\chi_k}\otimes \ket{\chi_{n\ominus_k}}\; \chi_{n}(x)^\star = \nonumber \\
&= \frac{1}{\sqrt{L}^2}\sum_{n=-\omega}^{+\omega} \sum_{k=-\omega}^{+\omega} \ket{\chi_k} \otimes\ket{\chi_{n\ominus_k}}\; \chi_{k}(x)^\star \chi_{n\ominus k}(x)^\star = \nonumber \\
&= \Big[\frac{1}{\sqrt{L}}\sum_{n'=-\omega}^{+\omega} \chi_{n'}(x)^\star \ket{\chi_{n'}}\Big] \otimes \Big[\frac{1}{\sqrt{L}}\sum_{k=-\omega}^{+\omega}  \chi_{k}(x)^\star\ket{\chi_k}\Big]= \nonumber \\
&= \sqrt{L}\ket{\delta_x} \otimes \sqrt{L} \ket{\delta_x}.
\end{align}
\hfill$\square$

% \paragraph{Proof of Theorem \ref{thm_PositionGeneratorsMomenta}}
% By the very definition of $\XbwmultSym$.
% \hfill$\square$

\vspace{-0.4cm}

\paragraph{A first connection between complementarity and the uncertainty principle}
Mutual unbias, also known as \textit{complementarity}, is sufficient to provide a first, rough approximation of the position/momentum uncertainty principle: it can be used to show that measuring any state first in the position observable and then in the momentum observable (or vice versa) always yields the totally mixed state. We sketch an informal, non-diagrammatic proof here, and leave a rigorous one (which uses Hopf law in the CPM category $\CPMCategory{\starHilbCategory}$) to future work on the topic. 

\noindent If one measures a pure state $\ket{\psi}$ in the position observable, one obtains a mixture $\rho$ of position eigenstates: $\rho := \int_{S^1} Q_x \ket{\psi}\bra{\psi} Q_x^\dagger \;dx = \int_{S^1} |\psi(x)|^2 \; \ket{\delta_x}\bra{\delta_x} \;dx$.
If one then measures the mixture $\rho$ in the momentum observable, one obtains a new mixture $\rho'$ given as follows:$\rho' := \sum_{n=-\omega}^{+\omega} P_{n\hbar} \rho P_{n\hbar}^\dagger = \sum_{n=-\omega}^{+\omega}\int_{S^1} |\psi(x)|^2 \; P_{n\hbar} \ket{\delta_x}\bra{\delta_x} P_{n\hbar}^\dagger \;dx$.
But $\rho$ is a classical mixture of position eigenstates, and mutual unbias alone is sufficient to conclude that $\rho'$ is the totally mixed state.
\hfill$\square$

\paragraph{Proof of Theorem \ref{thm_PositionObservable}}
The projector $Q_x$ is meant to send a standard state $\ket{\truncate{f}}$ of $\Ltwo{\reals/(L\integers)}$ to $f(x)$ times the (normalised) position eigenstate $\frac{\sqrt{L}}{\sqrt{2\omega + 1}}\ket{\delta_x}$ (times a normalisation factor making $Q_x$ idempotent). We expand a standard $f$ in the orthonormal basis of (normalised) momentum eigenstates as $\ket{\truncate{f}} = \sum_{n=-\omega}^{+\omega} \truncate{f}_n \frac{1}{\sqrt{L}}\ket{\chi_n}$, and we get:
\begin{align}
\Big(\id{} \otimes \frac{\sqrt{L}}{\sqrt{2\omega + 1}}\bra{\delta_x}\Big) \cdot \hbox{\begin{tikzpicture} [scale=0.8,transform shape] %% DO NOT CHANGE

\def\deltax{0.3} %% CAN BE CHANGED
\def\deltay{0.5} %% DO NOT CHANGE

%\path[use as bounding box] (-\deltax,-\deltay) rectangle (\deltax,\deltay);

\node (mult_label_outl) at (-\deltax,+\deltay) {};
\node (mult_label_outr) at (+\deltax,+\deltay) {};
\node [dot, fill=\Dcolour] (mult) at (0,0) {};
\node (mult_label_in) at (0,-\deltay) {};
\draw[-] [in=270,out=135] (mult) to (mult_label_outl);
\draw[-] [in=270,out=45] (mult) to (mult_label_outr);
\draw[-] (mult_label_in) to (mult);

%\draw (current bounding box.south west) rectangle (current bounding box.north east);
\end{tikzpicture}}\! \cdot \ket{f} 
&= \frac{\sqrt{L}}{\sqrt{2\omega + 1}}\sum_{n=-\omega}^{+\omega} \sum_{k = -\omega}^{\omega} \frac{1}{\sqrt{L}^2}\ket{\chi_k} \braket{\delta_x}{\chi_{n \ominus k}} \truncate{f}_n =\nonumber \\
&= \frac{\sqrt{L}}{\sqrt{2\omega + 1}}\sum_{n=-\omega}^{+\omega} \sum_{k = -\omega}^{\omega} \frac{1}{\sqrt{L}^2}\ket{\chi_k} \chi_{n \ominus k}(x) \truncate{f}_n = \nonumber \\
&= \frac{\sqrt{L}}{\sqrt{2\omega + 1}}\Big[ \sum_{k = -\omega}^{\omega} \frac{1}{\sqrt{L}}\chi_k(x)^\star\ket{\chi_k} \Big] \frac{1}{\sqrt{L}}\sum_{n=-\omega}^{+\omega}\chi_{n}(x) \truncate{f}_n = \nonumber \\
&= \frac{\sqrt{L}}{\sqrt{2\omega+1}}\ket{\delta_x} \; \frac{\sqrt{L}}{\sqrt{2\omega + 1}}\Big[\frac{1}{\sqrt{L}}\sum_{n=-\omega}^{+\omega}\chi_{n}(x)\frac{1}{\sqrt{L}^2}\braket{\chi_n}{\chi_n} \truncate{f}_n\Big] = \nonumber\\
&= \frac{\sqrt{L}}{\sqrt{2\omega+1}}\ket{\delta_x} \; \frac{\sqrt{L}}{\sqrt{2\omega + 1}}\braket{\delta_x}{\truncate{f}} \simeq \frac{\sqrt{L}}{\sqrt{2\omega + 1}}\big[f(x) \frac{\sqrt{L}}{\sqrt{2\omega + 1}} \ket{\delta_x}\Big].
\end{align}
\hfill$\square$

\vspace{-0.4cm}

\paragraph{Proof of Theorem \ref{thm_StrongComplementarity}}
Evaluating on momentum eigenstates, the three equations in  (\ref{eqn_StrongComplementarity}) are equivalent to the following equations, which are immediate to check:
\begin{align}
\Big(\frac{1}{\sqrt{L}}\ket{\chi_{n\oplus m}} \otimes \frac{1}{\sqrt{L}}\ket{\chi_{n \oplus m}}\Big) &= \hbox{\begin{tikzpicture} [scale=0.8,transform shape] %% DO NOT CHANGE

\def\deltax{0.3} %% CAN BE CHANGED
\def\deltay{0.5} %% DO NOT CHANGE

%\path[use as bounding box] (-\deltax,-\deltay) rectangle (\deltax,\deltay);

\node (mult_label_outl) at (-\deltax,+\deltay) {};
\node (mult_label_outr) at (+\deltax,+\deltay) {};
\node [dot, fill=\Zbwcolour] (mult) at (0,0) {};
\node (mult_label_in) at (0,-\deltay) {};
\draw[-] [in=270,out=135] (mult) to (mult_label_outl);
\draw[-] [in=270,out=45] (mult) to (mult_label_outr);
\draw[-] (mult_label_in) to (mult);

%\draw (current bounding box.south west) rectangle (current bounding box.north east);
\end{tikzpicture}}\! \frac{1}{\sqrt{L}}\ket{\chi_{n \oplus m}}\nonumber \\
\hbox{\begin{tikzpicture} [scale=0.8,transform shape] %% DO NOT CHANGE

\def\deltax{0.3} %% CAN BE CHANGED
\def\deltay{0.5} %% DO NOT CHANGE

%\path[use as bounding box] (-\deltax,-\deltay) rectangle (\deltax,\deltay);

\node (mult_label_outl) at (-\deltax,+\deltay) {};
\node (mult_label_outr) at (+\deltax,+\deltay) {};
\node [dot, fill=\Zbwcolour] (mult) at (0,0) {};
\node (mult_label_in) at (0,-\deltay) {};
\draw[-] [in=270,out=135] (mult) to (mult_label_outl);
\draw[-] [in=270,out=45] (mult) to (mult_label_outr);
\draw[-] (mult_label_in) to (mult);

%\draw (current bounding box.south west) rectangle (current bounding box.north east);
\end{tikzpicture}}\! \frac{1}{\sqrt{L}}\ket{\chi_{0}} &= (\frac{1}{\sqrt{L}}\ket{\chi_{0}} \otimes \frac{1}{\sqrt{L}}\ket{\chi_{0}}) \nonumber \\
\sqrt{L} \frac{1}{\sqrt{L}} \braket{\delta_0}{\chi_{n \oplus m}} &= \sqrt{L}^2 \frac{1}{\sqrt{L}^2}\braket{\delta_0}{\chi_{n}} \braket{\delta_0}{\chi_{m}}
\end{align}
Equation (\ref{eqn_WeylCCRs}) follows from strong complementarity, as shown in \cite{gogioso2015categorical}:
\begin{equation}
	\begin{multlined}
	\begin{tikzpicture}[node distance = 1cm]
		% Leftmost
		\node (center) {};
		\node[dot, fill=\Xbwcolour] (b) [below of = center, yshift = 0.5cm]{};
		\node (in) [below of = b,yshift = -0.25cm] {};
		\node[kpoint] (chi) [below left of = b] {$\chi_n$};
		\node (norm) [left of = chi, xshift = 2mm] {$\frac{1}{\sqrt{L}}$};
		\node[dot, fill=\Zbwcolour] (t) [above of = center, yshift = -0.5cm]{};
		\node (out) [above of = t,yshift = 0.25cm] {};
		\node[copoint] (delta) [above left of = t] {$\delta_x$};
		\node (norm) [left of = delta, xshift = 2mm] {$\sqrt{L}$};
		\begin{pgfonlayer}{background}
		\draw[-] (in) to (out);
		\draw[-,out=90,in=225] (chi) to (b);
		\draw[-,out=270,in=135] (delta) to (t);
		\end{pgfonlayer}
		% Center
		\node (equals) [right of = center] {$=$};
		\node (center) [right of = equals, xshift = 0.75cm] {};
		\node[dot, fill=\Zbwcolour] (bl) [below left of = center] {}; 
		\node[dot, fill=\Zbwcolour] (br) [below right of = center] {};
		\node[dot, fill=\Xbwcolour] (tl) [above left of = center] {};
		\node[dot, fill=\Xbwcolour] (tr) [above right of = center] {};
		\node[kpoint] (blin) [below of = bl] {$\chi_n$};
		\node (norm) [left of = blin, xshift = 2mm] {$\frac{1}{\sqrt{L}}$};
		\node (brin) [below of = br] {};
		\node[copoint] (tlout) [above of = tl] {$\delta_x$};
		\node (norm) [left of = tlout, xshift = 2mm] {$\sqrt{L}$};
		\node (trout) [above of = tr] {};
		\begin{pgfonlayer}{background}
		\draw[-] (bl) to (tl);
		\draw[-] (bl) to (tr);
		\draw[-] (br) to (tl);
		\draw[-] (br) to (tr);
		\draw[-] (blin) to (bl);
		\draw[-] (brin) to (br);
		\draw[-] (tl) to (tlout);
		\draw[-] (tr) to (trout);
		\end{pgfonlayer}
		% Rightmost
		\node (equals) [right of = center, xshift = 0.75cm] {$=$};
		\node (center) [right of = equals, xshift = 2.75cm] {};
		\node[dot, fill=\Zbwcolour] (b) [below of = center, yshift = 0cm]{};
		\node (in) [below of = b, yshift = 0.25cm] {};
		\node[copoint] (chi) [above left of = b] {$\delta_x$};
		\node (norm) [left of = chi, xshift = 2mm] {$\sqrt{L}$};
		\node[dot, fill=\Xbwcolour] (t) [above of = center, yshift = -0cm]{};
		\node (out) [above of = t, yshift = -0.25cm] {};
		\node[kpoint] (delta) [below left of = t] {$\chi_n$};
		\node (norm) [left of = delta, xshift = 2mm] {$\frac{1}{\sqrt{L}}$};
		\begin{pgfonlayer}{background}
		\draw[-] (in) to (out);
		\draw[-,out=270,in=135] (chi) to (b);
		\draw[-,out=90,in=225] (delta) to (t);
		\end{pgfonlayer}
		\node (factor) [left of = center, xshift = -1.75cm]{};
		\node[copoint] [above of = factor, yshift = -0.5cm](delta) {$\delta_x$}; 
		\node[kpoint] [below of = factor, yshift = 0.5cm](chi) {$\chi_n$}; 		
		\begin{pgfonlayer}{background}
		\draw[-] (chi) to (delta);
		\end{pgfonlayer}
	\end{tikzpicture}
	\end{multlined}
\end{equation}
\hfill$\square$

\end{document}